\documentclass[manuscript, screen, table]{acmart}

\usepackage{algorithmic}
\usepackage{textcomp}
\usepackage{booktabs}    % 专业的表格线条
\usepackage{tabularx}    % 支持自适应宽度的 X 列
\usepackage{array}       % 增强列格式控制
\usepackage{makecell}    % 支持表格内换行和多行文本
\usepackage{graphicx}    % 插入图片
\usepackage{subcaption}  % 支持子图排版

\def\BibTeX{{\rm B\kern-.05em{\sc i\kern-.025em b}\kern-.08em
    T\kern-.1667em\lower.7ex\hbox{E}\kern-.125emX}}

\setcopyright{none}

\begin{document}

% % 1. 干净的标题
% \title{Collaborative Air-Ground Sensing, Communication, Computing, Storage, and Intelligence for Low-Altitude Economy}
% 在 \title 后面紧跟一个方括号，填入缩写版的标题
\title[Collaborative Air-Ground SCCSI for LAE]{Collaborative Air-Ground Sensing, Communication, Computing, Storage, and Intelligence for Low-Altitude Economy}

% 2. 每位作者必须独立写明 \author, \email 和 \affiliation
\author{Yiqin Deng}
\email{yiqindeng@ln.edu.hk}
\affiliation{%
  \department{School of Data Science}
  \institution{Lingnan University}
  \city{Hong Kong}
  \country{China}
}

\author{Junhui Gao}
\email{junhui.gao@cityu.edu.hk}
\affiliation{%
 \department{Hong Kong JC STEM Lab of Smart City}
  \department{Department of Computer Science}
  \institution{City University of Hong Kong}
  \city{Hong Kong}
  \country{China}
}

\author{Zihan Fang}
\email{zihanfang3-c@my.cityu.edu.hk}
\affiliation{%
 \department{Hong Kong JC STEM Lab of Smart City}
  \department{Department of Computer Science}
  \institution{City University of Hong Kong}
  \city{Hong Kong}
  \country{China}
}

\author{Yanan Ma}
\email{yananma8-c@my.cityu.edu.hk}
\affiliation{%
 \department{Hong Kong JC STEM Lab of Smart City}
  \department{Department of Computer Science}
  \institution{City University of Hong Kong}
  \city{Hong Kong}
  \country{China}
}

\author{Xianhao Chen}
\email{xcheneee@hku.hk}
\affiliation{%
  \department{Department of Electrical and Electronic Engineering}
  \institution{The University of Hong Kong}
  \city{Hong Kong}
  \country{China}
}

\author{Yuguang Fang}
\email{my.fang@cityu.edu.hk}
\affiliation{%
  \department{Hong Kong JC STEM Lab of Smart City}
  \department{Department of Computer Science}
  \institution{City University of Hong Kong}
  \city{Hong Kong}
  \country{China}
}

\renewcommand{\shortauthors}{Y. Deng et al.}
% 3. 摘要必须在 \maketitle 之前
\begin{abstract}
Low-altitude economy (LAE) is transforming low-altitude airspace into a new cyber-physical infrastructure. Although air-ground communications have been widely studied, LAE is fundamentally different in the sense that it is mission-centric with diverse requirements, such as stringent safety and compliance constraints not be effectively addressed with a communication-centric design alone, which makes air-ground collaboration indispensable: Only through effectively coordinating air-ground infrastructure and resources can LAE missions be fulfilled. Consequently, LAE calls for task-driven, closed-loop, multi-resource orchestration of Sensing, Communication, Computing, Storage, and Intelligence (SCCSI), where key decisions must be co-designed under mobility and uncertainty. In this paper, we first present a novel framework that connects (i) LAE scenarios and a requirement--resource coupling matrix, (ii) an air--ground collaborative architecture, and (iii) methodological toolboxes for SCCSI co-optimization and online decision-making. We then systematically review enabling technologies for collaborative SCCSI resources and capabilities, emphasizing their coupling and end-to-end tradeoffs. Finally, we summarize testbeds, datasets, and evaluation metrics, and provide representative use cases to illustrate how the proposed framework translates application requirements into practical task-driven optimization designs, together with open challenges and a roadmap toward scalable and trustworthy LAE deployment.
\end{abstract}

% 4. ACM 专属的关键字格式，替换掉了 IEEEkeywords，必须放在 \maketitle 之前

\authorsaddresses{%
  % 手动编写页脚内容，可以把同一单位的作者写在一起
Yiqin Deng is with Lingnan University, Hong Kong, China (email: yiqindeng@ln.edu.hk);
Junhui Gao, Zihan Fang, Yanan Ma, and Yuguang Fang are with City University of Hong Kong, Hong Kong, China (emails: junhui.gao@cityu.edu.hk, \{zihanfang3-c, yananma8-c\}@my.cityu.edu.hk, my.Fang@cityu.edu.hk); 
Xianhao Chen is with the University of Hong Kong, Hong Kong, China (email: xcheneee@hku.hk).
}
% 5. 生成标题和作者块
\maketitle

% 正文从这里开始
\section{Introduction}
Low-altitude economy (LAE) refers to economic activities in the airspace under $1,000$ meters above ground (with extensions to $3,000$ meters for specialized applications), enabled by advanced air mobility platforms such as unmanned aerial vehicles (UAVs\footnote{In this paper, the terms ``UAV'' and ``drone'' are used interchangeably, referring to the same type of unmanned aerial vehicle.}), electric vertical take-off and landing (eVTOL) aircraft, and air taxis. These technologies are catalyzing transformative applications in urban logistics, precision agriculture, emergency medical supply delivery, intelligent transportation management and control, infrastructure inspection, and tourism~\cite{Policy2024,Jiang2025integrated}.

\subsection{From application-driven requirements to air--ground collaboration framework}
A distinctive feature of LAE is its \emph{application-driven heterogeneity}. Different LAE missions impose vastly different and often stringent requirements on latency, reliability, safety/compliance, coverage, energy efficiency, and situational awareness. Meanwhile, the LAE operating environment is highly dynamic due to fast-varying 3D mobility, frequent topology changes, and massive spatiotemporal data generation. Similar heterogeneity has been observed in UAV-assisted IoT systems, where diverse application classes exhibit distinct QoS requirement patterns (e.g., latency, reliability, and coverage), thereby motivating task-aware system design~\cite{adil2024uav}. These factors collectively motivate a fundamental paradigm shift: LAE systems cannot be effectively engineered by optimizing a single resource or a single layer in isolation. Instead, they must be designed as \emph{mission-centric cyber-physical systems} where sensing, communication, computing, storage, and intelligence are jointly orchestrated in a closed loop.

\begin{figure*}[tb!]
    \centering
    \includegraphics[width=0.825\textwidth]{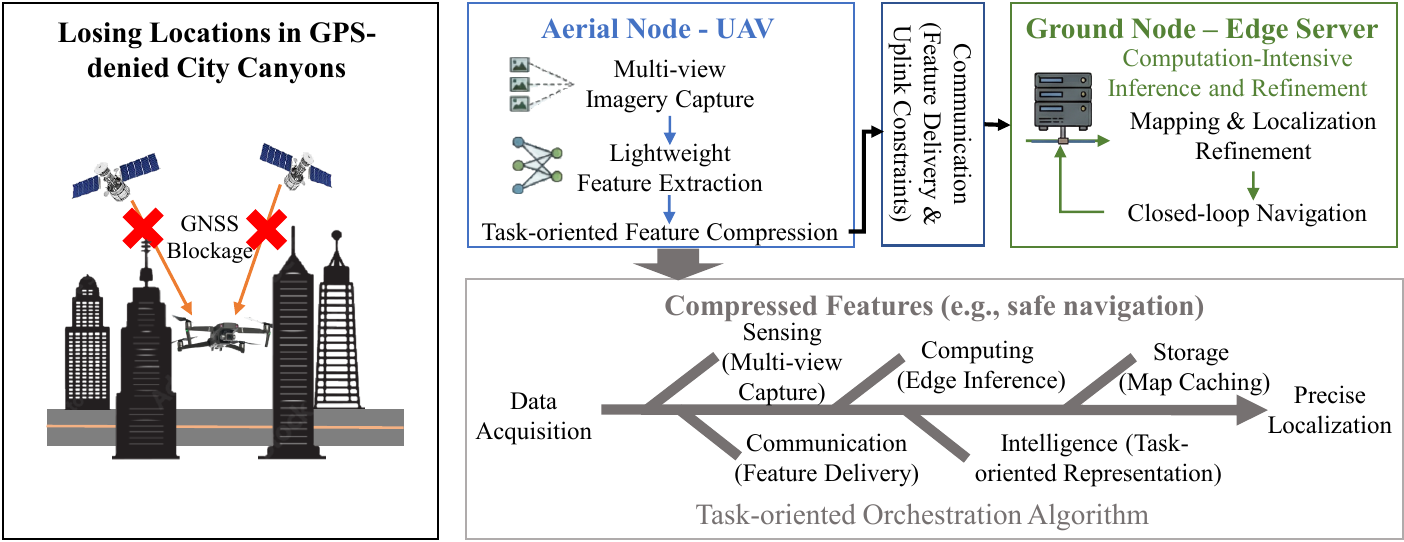}
    \caption{\small{A motivating example of a mission-driven LAE task: high-precision UAV navigation in GNSS-denied urban canyons via air--ground collaborative Sensing, Communication, Computing, Storage, and Intelligence (SCCSI) orchestration. The UAV (air side) performs lightweight multi-view sensing and task-oriented feature compression, transmits compact representations over air--ground links, and a ground edge server (ground side) executes computation-intensive inference and map/model-assisted refinement. The resulting navigation state is fed back to the UAV for closed-loop control, illustrating task-driven co-design across \emph{Sensing, Communication, Computing, Storage, and Intelligence} over the air-and-ground dimensions.}}
    \label{fig:example}
\end{figure*}

Crucially, the above application-driven requirements \emph{necessitate} \emph{air--ground collaboration}. Aerial platforms provide unique OTA advantages, rapid deployment, flexible 3D mobility, line-of-sight (LoS) opportunities, and bird's-eye-view (BEV) sensing, which are invaluable for coverage extension and wide-area perception. However, UAVs are inherently constrained by limited battery energy, payload, and onboard computing/storage constraints; simply ``pushing more resources onboard'' (e.g., adding GPUs) increases weight and power consumption and is often not practical. In contrast, ground infrastructures (e.g., 5G/6G base stations, cloud/edge computing, and intelligent transportation systems) offer a stable power supply, stronger computing/storage capabilities, and persistent connectivity, yet they often face severe line-of-sight (LoS) obstructions and multipath interference within urban environments, resulting in restricted aerial coverage and awareness. Therefore, LAE service provisioning naturally becomes a two-dimensional (air-and-ground) system design problem: the effective solution is not more resources in one domain, but better orchestration of resources across both domains.

\subsection{Motivating example: task-driven air--ground resource synergy}

To better motivate the aforementioned thinking, we present a motivating example for high-precision and reliable navigation in GNSS-denied urban canyons, where GNSS signals can be severely impaired by blockage and multipath. More broadly, high-performance positioning, navigation, and timing (PNT) is a cornerstone for safe low-altitude operations, and recent studies have discussed space--air--ground collaborative PNT architectures and enabling technologies for LAE~\cite{mingquan2025space}.

As shown in Fig.~\ref{fig:example}, in urban canyons, GNSS can fail, while standalone onboard solutions, such as IMU and visual odometry, typically suffer from drift accumulation and substantial real-time computation demands. Although multi-view imagery can improve spatial awareness and localization accuracy, lightweight UAVs often cannot afford high-dimensional visual processing due to computing and energy constraints. To address this challenge, our prior work~\cite{Fang2025Task} demonstrated an edge--aerial collaboration paradigm, where a UAV performs lightweight sensing and task-oriented feature extraction/compression, and the ground edge server executes computation-intensive inference and map/model-assisted refinement. This closed-loop workflow enables reliable navigation under stringent uplink and latency constraints.

This task-driven design illustrates that a single LAE mission, such as localization and navigation, may inherently require coordinated \emph{sensing} for multi-view capture, \emph{communication} for feature delivery, \emph{computing} for edge inference, \emph{storage} for localized map caching and data buffering, and \emph{intelligence} for task-oriented representation and decision making. It also reflects a broader trend: as LAE scales to dense operations, such as logistics swarms, emergency response, and airspace monitoring, SCCSI resources must be jointly managed across air and ground under safety and reliability constraints. This motivates task-driven orchestration algorithms rather than siloed optimization of individual resources.

\subsection{Motivations of AG-SCCSI: a task-driven, closed-loop orchestration problem}
Motivated by the above observations, we focus on Air--Ground integrated Sensing, Communication, Computing, Storage, and Intelligence (AG-SCCSI) as a unified paradigm for LAE. Applications such as urban air mobility, emergency rescue, and large-scale drone logistics require ultra-low latency, ultra-high reliability, and pervasive availability of SCCSI resources across vast, dynamic three-dimensional (3D) airspace~\cite{Zeng2024Accessing}. Moreover, safety is a first-order requirement for dense low-altitude operations. Systematic evidence on UAV operational safety assessment further underscores the need for safety-aware orchestration and verifiable decision pipelines~\cite{asghari2025uav}. 

However, neither aerial platforms nor terrestrial systems alone can sustainably meet these mission-level QoS and safety requirements. UAVs excel at flexible sensing and agile connectivity but remain energy- and payload-constrained with volatile links under mobility~\cite{Fan2024Channel}. Terrestrial systems provide stronger and steadier compute/storage and more stable connectivity, but face limited aerial coverage and reduced responsiveness to aerial dynamics~\cite{Jiang2025integrated}. This complementarity suggests a fundamental design principle:
\begin{quote}
\emph{Application-driven LAE requirements necessitate air--ground collaboration, which in turn requires task-driven, closed-loop orchestration of SCCSI resources across the air-and-ground dimensions.}
\end{quote}

Unfortunately, this design also poses several system-level challenges:
\begin{itemize}
    \item \textit{Resource heterogeneity and abstraction}: Air and ground nodes differ in mobility, energy models, compute/storage budgets, sensing modalities, and communication interfaces. Effective collaboration requires unified abstractions and interoperable orchestration interfaces.
    \item \textit{Mobility-induced volatility}: UAV mobility causes rapid topology changes and fluctuating resource availability, demanding online association, adaptive service migration, and resilient task offloading.
    \item \textit{Coupled trade-offs across SCCSI}: Offloading reduces onboard energy consumption but incurs communication overhead and handover costs; caching reduces latency but competes with onboard storage; sensing/reporting improves situational awareness but increases communications and computing workloads. These coupling effects require joint, task-driven optimization.
    \item \textit{Security, safety, and compliance}: Hybrid air--ground operation introduces threats such as spoofing, jamming, and data poisoning, and mandates trustworthy coordination mechanisms and auditable data/decision pipelines.
    \item \textit{Scalability in dense operations}: Managing thousands of aerial/ground agents in urban airspace requires scalable, lightweight, and distributed orchestration architectures.
\end{itemize}

Existing air--ground integration often treats UAVs as passive relays or views ground infrastructure as a simple backhaul for communication and offloading. Such designs underutilize the mutual complementarity. In LAE, UAV BEV sensing can easily augment ground perception in occluded environments, while ground edge/cloud can sustain intensive computing and storage (e.g., map/model/service caching and compliance logging). These bidirectional synergies motivate AG-SCCSI as a unified framework, enabling both \emph{vertical collaboration} (e.g., air--ground task offloading, ground-assisted navigation, and OTA advantages)) and \emph{horizontal collaboration} (e.g., UAV swarm sensing, edge federation among ground nodes), ultimately providing adaptive and mission-tailored service provisioning for LAE~\cite{Liu2024Sensing,deng2024uav_wcm,Deng2024UAV}.

\subsection{Related Works and Our Contributions}
Before presenting our AG-SCCSI framework, we compare our work with existing surveys and vision papers. Recent papers have advocated low-altitude aerial intelligent networks (LAIN) to support large-scale LAE services. However, how to translate heterogeneous mission requirements into task-driven, closed-loop orchestration across SCCSI resources remains underexplored~\cite{yuhong2025towards}.

\begin{itemize}
    
 \item \textit{Space-Air-Ground Integrated Networks (SAGIN)}: Research in this area, such as Xiao \textit{et al.}~\cite{xiao2024space}, often explores broad architectural paradigms for global connectivity, leveraging satellites, aerial platforms, and terrestrial networks for communication. While SAGIN aims for ubiquitous communication coverage, its primary focus is to address the limitations of single networks  by integrating heterogeneous segments (space, air, and ground) that often operate on different communication media and network protocols, leading to a ``bucket effect'' where overall system capability is limited by the weakest part. This body of work, including discussions on SAGIN integration by Wang \textit{et al.}~\cite{wang2025bridging}, provides a foundational vision on integrated networks. However, its scope is typically too macroscopic, emphasizing extensive communication coverage and traffic sharing/offloading across vast geographical scales, often overlooking the unique, fine-grained microscopic demands of agile and resource-constrained applications, such as urban drone logistics or emergency response in LAE. Thus, SAGIN alone cannot address the challenges in LAE where coordinated AG-SCCSI resources are needed. 

\item \textit{Intelligent Network Paradigms and Semantic Communications}: Another strand of research focuses on infusing intelligence into communication networks. Examples include integrating advanced AI models (e.g., large language models and diffusion models) for intelligent, task-driven networking in SAGIN~\cite{Gao2025Space}, the use of generative AI to assist network automation and management~\cite{zhang2024generative}, and emerging paradigms of semantic communication that transmit semantic contexts instead of raw bits~\cite{meng2025semantics}. These works promote task-oriented, AI-driven networking to enhance efficiency and adaptability. Nevertheless, their discussions remain mostly either conceptual or high-level, focusing on general automation or computing improvements. They have not explicated how ``intelligence'' can be tightly coordinated with sensing, communication, and computing resources under the dynamic conditions of LAE. In particular, the unique challenges of LAE (e.g., high mobility, strict resource constraints, cross-domain coordination, (near) real-time missions) are not directly addressed by these works.

\item \textit{Integrated Sensing and Edge AI (ISEA)}: After collecting data, intelligence at the edge can be extracted immediately. Thus, ISEA was proposed and investigated lately. Notably, Jiang \textit{et al.} introduced ISEA as a new paradigm for 6G intelligent perception~\cite{Jiang2025ISEA}. ISEA deeply integrates the acquisition, processing, and transmission of sensing data with edge AI, featuring a holistic co-design of sensing, communication, and computation for intelligence. This represents a significant step toward functional convergence of sensing, communication, and computing. However, ISEA is geared toward general 6G use-cases and lacks LAE-specific design considerations. It has not fully explored the tight integration of all five SCCSI dimensions, notably missing explicit roles of storage (e.g., caching for computation offload) and distributed intelligence (e.g., federated learning across air/ground nodes), under high mobility and stringent resource constraints of LAE missions.

\item \textit{Low-Altitude Wireless Networks (LAWN) and LAE-Native Network Architectures:}
Recent works have started to examine LAE from a more native networking perspective, beyond generic UAV communications or broad SAGIN abstractions. In particular, Yuan \textit{et al.} introduce LAWNs as a reconfigurable three-dimensional layered architecture that tightly integrates connectivity (communications), sensing, control, and computing across aerial and terrestrial nodes. In parallel, Wu \textit{et al.} provide a comprehensive survey of LAWN systems, covering functional evolution, performance metrics, privacy and security, and airspace architecture. These works substantially enrich the architectural understanding of LAE-native networks. However, they mainly emphasize communication, control, and airspace-level integration, and have not explicitly formulated a mission-centric closed-loop orchestration framework that jointly coordinates SCCSI across air and ground nodes~\cite{lawn_ground_to_sky_2025}.

\item \textit{LAE Network Architecture, Standards, and Integrated Technologies:}
Recent studies have also started to examine LAE from the perspective of native network architecture and system realization. In particular, Wang~\textit{et al.}~\cite{wang2025bridging} discuss the core architecture, related standards, and integrated technologies of LAE networks, covering communication, sensing, computing, positioning, navigation and surveillance, flight control, and airspace management within a unified framework. Compared with generic UAV networking or broad SAGIN abstractions, this line of work provides a more direct system-level view of how LAE networks can be realized in practice. However, its main focus remains on architecture realization, standards support, and multi-technology integration, rather than on a mission-centric closed-loop orchestration framework that explicitly coordinates SCCSI across air and ground nodes~\cite{lae_realization_2025}.
\end{itemize}

\begin{table}[htbp] % 去掉星号，改为单栏标准 table
\caption{Comparison of Survey and Vision Papers on Air-Ground Integration and SCCSI-Related Technologies.}
\label{tab:survey_comparison}
\vspace{-0.5em}
\centering
\footnotesize
\renewcommand{\arraystretch}{1.25} % 稍微拉大行距，容纳堆叠的文字
\setlength{\tabcolsep}{4pt}        % 优化列间距

% 使用 tabularx：前两列自适应左对齐，后两列固定宽度并居中
\begin{tabularx}{\textwidth}{@{} 
  l 
  >{\hsize=0.9\hsize\raggedright\arraybackslash}X 
  >{\hsize=1.1\hsize\raggedright\arraybackslash}X 
  >{\centering\arraybackslash}p{2.5cm} 
  >{\centering\arraybackslash}p{2.2cm} 
  @{}}
\toprule
\textbf{Reference} & \textbf{Focus Area} & \textbf{Dimensions Covered} & \textbf{\makecell{Air-Ground\\Coordination?}} & \textbf{\makecell{LAE-\\Specific?}} \\
\midrule

\cite{xiao2024space} & LLM, Diffusion \& DT for SAGIN & Sens., Comm., Comp. & \makecell[c]{\ensuremath{\checkmark}\\ \scriptsize(Multi-layer)} & $\times$ \\ 
\addlinespace

\cite{zhang2024generative} & LLM-Driven SAGIN & Sens., Comm. & \makecell[c]{\ensuremath{\checkmark}\\ \scriptsize(Multi-layer)} & $\times$ \\ 
\addlinespace

\cite{wen2024survey} & ISCC in 6G & Sens., Comm., Comp. & \makecell[c]{\ensuremath{\checkmark}\\ \scriptsize(Implicit)} & $\times$ \\
\addlinespace

\cite{Zhen2024Air-ground} & AG-MEC for Task Offloading & Comm., Comp. & \ensuremath{\checkmark} & $\times$ \\
\addlinespace

\cite{wu2024ai} & AI-Driven Deployment Automation & Comm., Comp., Intell. & \ensuremath{\checkmark} & $\times$ \\
\addlinespace

\cite{Jiang2025ISEA} & ISEA in 6G & Sens., Comm., Comp., Intell. & \makecell[c]{\ensuremath{\checkmark}\\ \scriptsize(Implicit)} & $\times$ \\
\addlinespace

\cite{wang2025bridging} & SAGIN & Sens., Comm., Comp., Intell. & \makecell[c]{\ensuremath{\checkmark}\\ \scriptsize(General NTN)} \\
\addlinespace

\cite{Getu2024Semantic} & Semantic Communication for 6G & Sens., Comm., Intell. & \makecell[c]{\ensuremath{\checkmark}\\ \scriptsize(Implicit)} & $\times$ \\
\addlinespace

\cite{meng2025semantics} & Semantic-Enabled SAGIN & Sens., Comm., Comp., Intell. & \makecell[c]{\ensuremath{\checkmark}\\ \scriptsize(General NTN)} \\
\addlinespace

\cite{Telikani2025} & UAV-aided ITS & Sens. (Visual), Comm., Planning & \makecell[c]{\ensuremath{\checkmark}\\ \scriptsize(Sub-domain)} \\
\addlinespace

\cite{Jiang2025integrated} & ISAC for LAE & Sens., Comm. & \ensuremath{\checkmark} & \ensuremath{\checkmark} \\
\addlinespace

\cite{lawn_ground_to_sky_2025} & Low-altitude wireless networks & Sens., Comm., Comp. \newline \scriptsize(+Ctrl./ATM) & \ensuremath{\checkmark} & \ensuremath{\checkmark} \\
\addlinespace

\cite{lae_realization_2025} & LAE network architecture  & Sens., Comm., Comp., Intell. \newline \scriptsize(+PNT/ATM) & \ensuremath{\checkmark} & \ensuremath{\checkmark} \\
\addlinespace

\cite{Chen2024} & VaaS for Smart Cities & Sens., Comm., Comp., Storage, Intell. & $\times$ & $\times$ \\
\midrule
\textbf{This Work} & \textbf{AG-SCCSI for LAE} & \textbf{All SCCSI} & \textbf{\ensuremath{\checkmark}} & \textbf{\ensuremath{\checkmark}} \\
\bottomrule
\end{tabularx}
\end{table}

In summary, while numerous studies address subsets of the SCCSI dimensions (e.g., ISAC for sensing-communication, MEC for computing offload) or propose broad architectures and integrated technology roadmaps (e.g., SAGIN, LAWN, and LAE-native network realization frameworks) for next-generation networks, there is a persistent gap in the literature: no existing work provides a holistic survey or framework that systematically defines and integrates all five SCCSI dimensions in a unified air-ground paradigm tailored for LAE. This gap is critical, since LAE’s distinctive requirements for agility, resource efficiency, timely decision making, and cross-domain coordination can only be met with a mission-critical holistic framework that transcends generic architectures or partial integrations.

To our best knowledge, \textit{this is the first survey} to systematically define and explore the AG-SCCSI framework as a unified paradigm specifically for LAE. To visually underscore the novelty and comprehensiveness of our work, we provide a detailed comparison with existing survey papers in Table~\ref{tab:survey_comparison}. 
Our main contributions are summarized as follows. 
\begin{itemize}
\item \textit{Holistic AG-SCCSI Framework:} We propose a novel AG-SCCSI framework, which cohesively unifies multi-dimensional SCCSI resources in 3D space across diverse air and ground platforms. Unlike prior architectures like SAGIN (focused on communication connectivity scale) or ISEA (focused on sensing–AI fusion), AG-SCCSI is explicitly designed for LAE, enabling seamless and efficient coordination of all five SCCSI resource dimensions. This framework directly addresses the interplay among sensing data, wireless links, computing tasks, cached contents, and intelligent navigation and control in air–ground networks.
\item \textit{Comprehensive Survey of Enabling Technologies:} We conduct a thorough survey of key enabling technologies that underpin AG-SCCSI, including Integrated Sensing and Communication (ISAC), Air–Ground Mobile Edge Computing (AG-MEC), Digital Twin (DT) modeling, Federated Learning (FL), and Semantic Communications. Our analysis goes beyond generic overviews by focusing on each technology’s specific role and synergies in air–ground integration for LAE. We highlight how these technologies interconnect within the AG-SCCSI framework, providing insights not covered in existing surveys that consider them mostly in isolation.
\item \textit{Concrete LAE Application Scenarios and Validation:} We present multiple application scenarios in the LAE application domains, such as emergency response, intelligent traffic management, and urban drone delivery logistics, and analyze how the AG-SCCSI framework elevates service capabilities and operational efficiency in each case. Through these examples, we illustrate the practical benefits of tightly coordinated SCCSI resources (e.g., faster situational awareness, improved reliability, real-time decision making). Furthermore, we discuss potential validation methodologies (e.g., prototyping, simulation testbeds, and field trials) for these scenarios, bridging the gap between theoretical framework and real-world deployment.
\item \textit{Future Research Directions:} We identify several open challenges and promising future research directions to advance AG-SCCSI in LAE. These include developing semantic communication techniques for efficient mission-critical information exchange, exploring LLM-driven UAV swarm control for enhanced autonomy, and innovating sustainable energy management to prolong aerial operations. By highlighting these pressing issues and opportunities, we aim to guide and inspire future research efforts in building intelligent, robust, and sustainable LAE systems.
\end{itemize}
%A comparison with existing survey papers, highlighting the novelty and comprehensiveness of our work, is provided in Table~\ref{tab:survey_comparison}.

% ===============
% LIST OF ABBREVIATIONS (Optimized for single-column ACM format)
% ===============
\begin{table}[t!]
\centering
\caption{List of Abbreviations}
\label{tab:abbreviations}
\vspace{-0.5em}
\footnotesize
\setlength{\tabcolsep}{6pt}
\renewcommand{\arraystretch}{1.1}

\begin{tabularx}{\textwidth}{@{} l >{\raggedright\arraybackslash}X l >{\raggedright\arraybackslash}X @{}}
\toprule
\textbf{Abbreviation} & \textbf{Description} & \textbf{Abbreviation} & \textbf{Description} \\
\midrule
ABS      & Aerial Base Station & ADMM     & Alternating Direction Method of Multipliers \\
AG-MEC   & Air-Ground Multi-access Edge Computing & AG-SCCSI & Air-Ground Sensing, Communication, Computing, Storage \& Intelligence \\
AirComp  & Over-the-Air Computation & AoD      & Angle of Departure \\
AoI      & Age of Information & AP       & Access Point \\
BCD      & Block Coordinate Descent & BEV      & Bird's-eye View \\
BS       & Base Station & CDT      & Cooperative Digital Twin \\
CEN      & Cache-Enabled Network & CLP      & Cooperative Long-Term Average Optimization \\
C-SLAM   & Cooperative Simultaneous Localization and Mapping & CPN      & Computing Power Network \\
DRL      & Deep Reinforcement Learning & DT       & Digital Twin \\
DSA      & Dynamic Spectrum Access & ES       & Edge Server \\
FL       & Federated Learning & FMCW     & Frequency-Modulated Continuous Wave \\
GNSS     & Global Navigation Satellite System & GPS      & Global Positioning System \\
GV       & Ground Vehicle & HAP      & High-Altitude Platform \\
HFL      & Hierarchical Federated Learning & HFRL     & Heterogeneous Federated Reinforcement Learning \\
IMU      & Inertial Measurement Unit & ISAC     & Integrated Sensing and Communication \\
ISCC     & Integrated Sensing--Communication--Computing & ITS      & Intelligent Transportation Systems \\
KG       & Knowledge Graph & KKT      & Karush-Kuhn-Tucker Conditions \\
LAE      & Low-Altitude Economy & LiDAR    & Light Detection and Ranging \\
LoS      & Line of Sight & MAB      & Multi-Armed Bandit \\
MAPPO    & Multi-Agent Proximal Policy Optimization & MARL     & Multi-Agent Reinforcement Learning \\
MCTS     & Monte Carlo Tree Search & MEC      & Multi-access Edge Computing \\
mD       & Multi-Dimensional (links/metrics) & NLoS     & Non-Line of Sight \\
OTA      & Over-the-Air & OTA-FL   & Over-the-Air Federated Learning \\
PLS      & Physical-Layer Security & PPO      & Proximal Policy Optimization \\
PS       & Parameter Server & QoE      & Quality of Experience \\
QoS      & Quality of Service & RF       & Radio Frequency \\
RIS      & Reconfigurable Intelligent Surface & RSU      & Roadside Unit \\
SA       & Semantic Attention & SAoI     & Semantic-Aware Age of Information \\
SAGIN    & Space-Air-Ground Integrated Network & SCA      & Successive Convex Approximation \\
SDN      & Software-Defined Networking & SemCom   & Semantic Communication \\
SLP      & Symbol-Level Precoding & SNR      & Signal-to-Noise Ratio \\
STAR-RIS & Simultaneously Transmitting \& Reflecting RIS & UAS      & Unmanned Aircraft System \\
UAV      & Unmanned Aerial Vehicle & UWB      & Ultra-Wideband \\
VPS      & Visual Positioning System & VSP      & Virtual Service Provider \\
\bottomrule
\end{tabularx}
\end{table}

\subsection{Organization}
The rest of this paper is organized as follows. Section~2 introduces the AG-SCCSI framework and the associated air-ground resource orchestration mechanisms. Section~3 surveys the enabling technologies that support integrated AG-SCCSI in LAE. Section~4 reviews representative datasets, testbeds, digital twins emulation, simulators, and evaluation metrics for AG-SCCSI validation. Section~5 presents representative application scenarios. Section~6 discusses open challenges and future research directions. Section~7 concludes the paper. For convenience, Table~\ref{tab:abbreviations} summarizes the abbreviations used throughout the paper.

\section{AG-SCCSI Framework and Resource Orchestration Mechanisms}
\label{section: framework}

In this section, we present the AG-SCCSI framework and the main resource orchestration mechanisms for LAE applications. We emphasize the \emph{vertical collaborations} across air and ground for each SCCSI element and the corresponding orchestration mechanisms that together constitute our AG-SCCSI framework.

\begin{figure*}[tb!]
    \centering
    \includegraphics[width=0.925\textwidth]{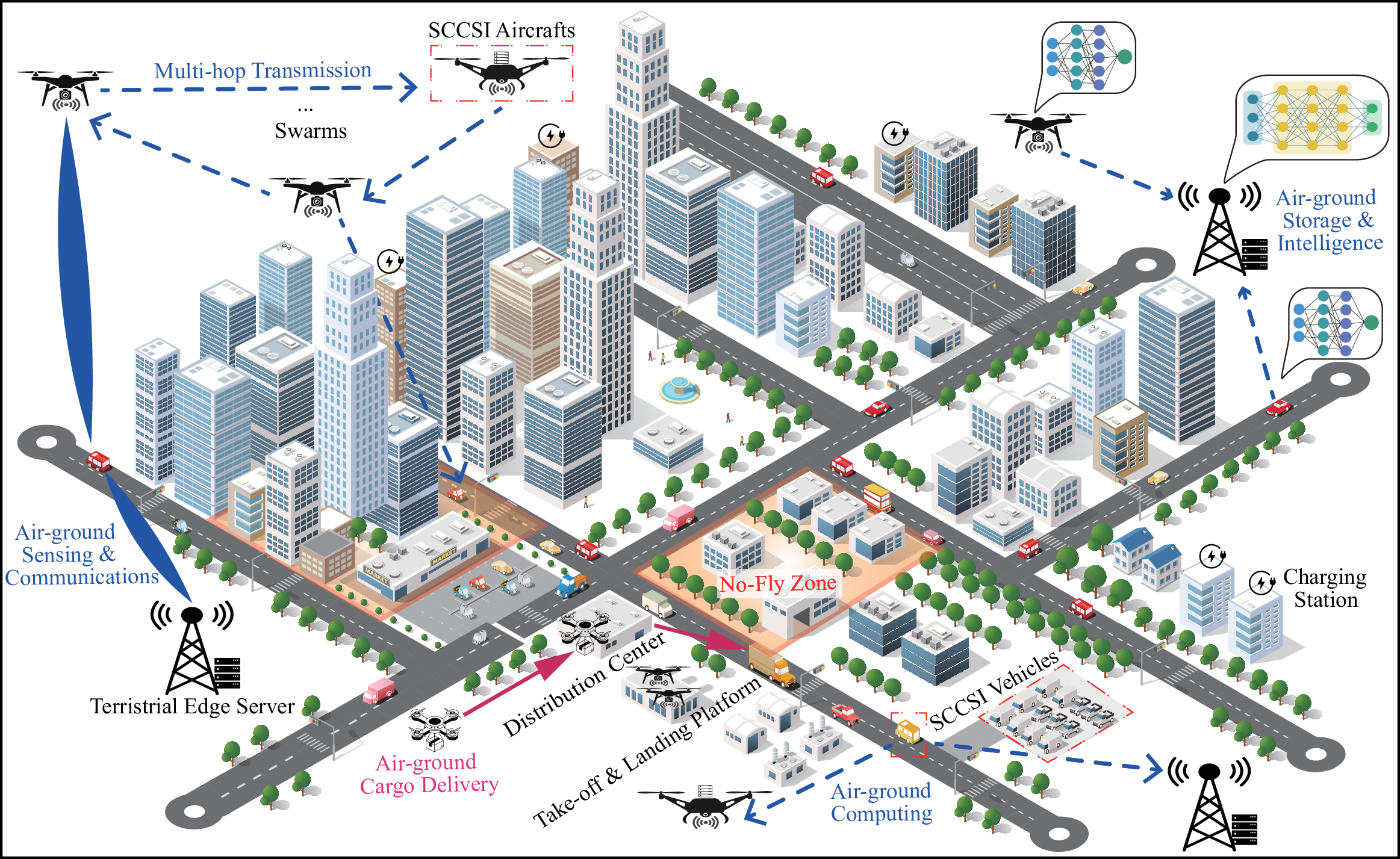}
    \caption{\small{Scenario illustration of collaborative air-ground sensing, communication, computing, storage, and intelligence in the low-altitude economy, featuring no-fly zones, charging stations, and other key elements, along with various applications requiring different air-ground SCCSI resources.} }
    \label{fig:architecture}
\end{figure*}

\subsection{AG-SCCSI Framework}
As shown in Fig.~\ref{fig:architecture}, the AG-SCCSI framework comprises UAVs, ground edge servers (ESs), base stations (BSs)/access points (APs)/roadside units (RSUs), and SCCSI-capable ground vehicles (GVs). UAVs serving LAE missions (e.g., parcel, medical, and food delivery) form OTA mobile networks, providing OTA sensing and communications, while ground nodes (ESs, BSs/APs/RSUs, and GVs) contribute ground sensing, communications, and computing. Together they compose a 3D service network with 3D SCCSI resources for LAE missions and smart-city operations and services~\cite{fang2025resources,Chen2024}. Analogous to ground SCCSI service networks \cite{fang2025resources,Chen2024}, mobile things carry SCCSI resources via application-specific \emph{Points of Connection (PoCs)\footnote{A PoC is a conceptual device with customized resources or capability for SCCSI services \cite{fang2025resources,Chen2024}.}}. Different from ground deployments~\cite{fang2025resources,Chen2024,ding2018smart,ding2017cognitive}, airborne platforms face strict payload and energy limitations, and hence PoC customization must make tradeoff between capability and weight/power, affecting airframe mechanics and mission endurance.

A second fundamental difference from legacy communication frameworks lies in \emph{supported services}: AG-SCCSI links are not only for message passing but for ultimate \emph{mission fulfillment}. For instance, OTA links may be established for OTA sensing to guide UAV navigation during delivery, or to discover remote ESs to complete offloaded computing tasks \cite{Deng2024UAV}. Consequently, AG-SCCSI must manage \emph{multi-dimensional (mD) links} between node pairs under heterogeneous physical and logical constraints, with objectives defined by mission semantics rather than pure data throughput. This 3D task-oriented perspective complements the macroscopic coverage-centric view of SAGIN and is essential in LAE where coordinated SCCSI is the bottleneck rather than mere connectivity \cite{xiao2024space}. 

\subsection{AG-SCCSI Network Formation and Management}

Beyond generic SAGIN formation \cite{xiao2024space}, AG-SCCSI forms \emph{mission-specific} topologies with mD link metrics. For example, for UAV navigation and perception, candidate links must jointly satisfy perception-quality and communication-quality constraints so that an ego UAV can select neighbors whose sensed views support robust fusion\footnote{This is analogous to collaborative vision fusion for vehicles on the ground, but UAV images are often offloaded to an ES on the ground for perception fusion due to onboard computing limitations.} \cite{Fang2025Task}. Hypergraph abstractions naturally capture such mD relationships. 

Given high mobility (e.g., for UAVs and GVs), fading, and opportunistic ground computing, the control plane must proactively learn network/resource states and perform low-latency decision making. We advocate a software-defined network (SDN)  design approach with an \emph{intent-driven} controller that fuses measurements and predictions (e.g., via a  \emph{digital twin}, DT) for online what-if evaluation and look-ahead orchestration. Emerging O-RAN DT studies and industry reports show DTs improve RAN monitoring, policy testing and closed-loop optimization, making them suitable for AG-SCCSI state tracking and ``plan–do–check–act'' control~\cite{nguyen2025dtoran,oran_dtran_2024}. 

\begin{figure*}[tb!]
    \centering
    \includegraphics[width=0.825\textwidth]{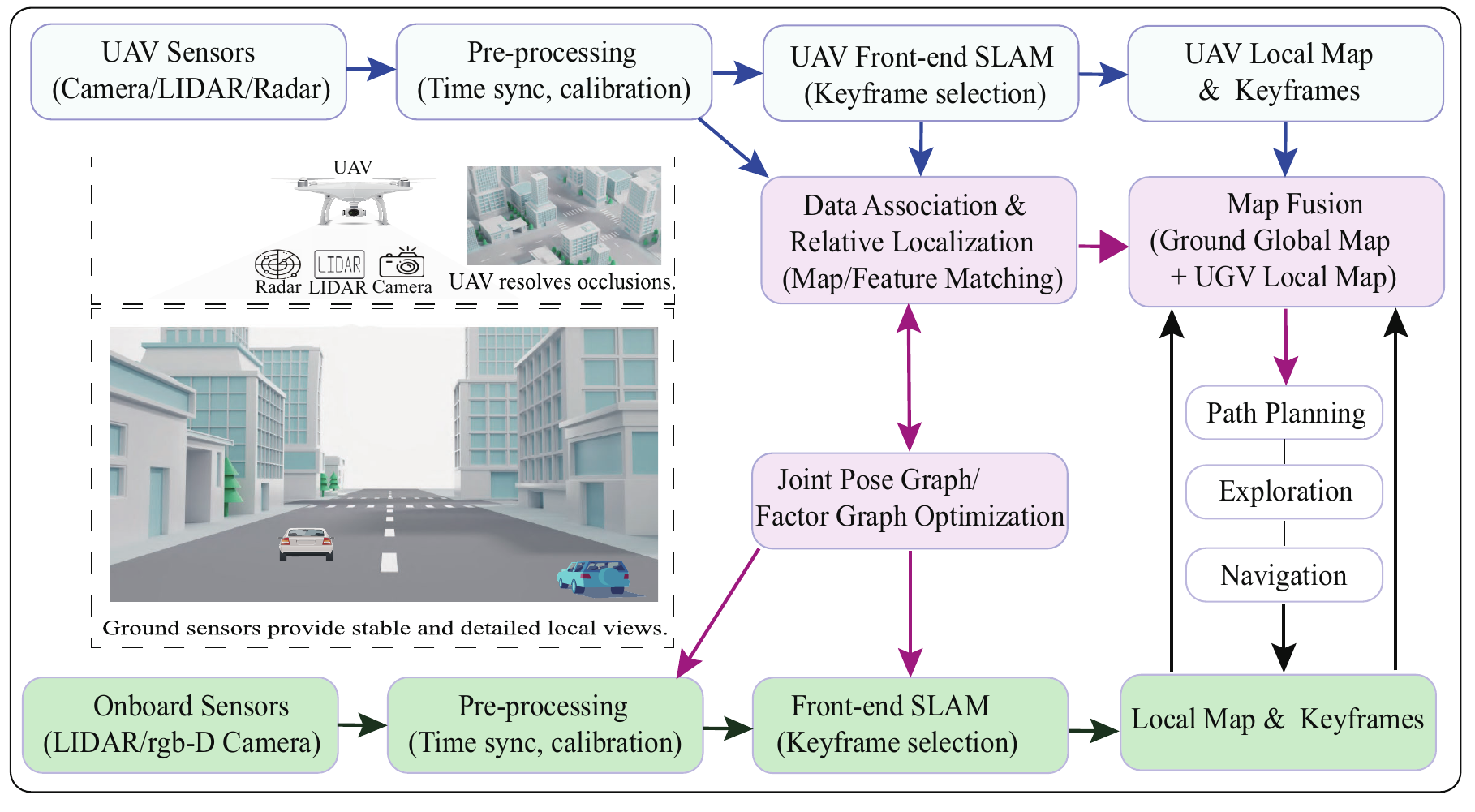}
    \caption{\small{Air-Ground Sensing Coordination:  Simultaneous Localization and Mapping } }
    \label{fig:sensing}
\end{figure*}

\subsection{Orchestrating Air--Ground Sensing}

Sensing in 3D urban environments benefits from the additional degrees of freedom (DoFs) offered by aerial platforms. Cooperative SLAM and air--ground collaboration (AGC) allow UAVs to complement ground robots or vehicles in occluded, congested, or otherwise unreachable areas. As illustrated in Fig.~\ref{fig:sensing}, a robust AGC system typically involves parallel processing pipelines, where both aerial and ground nodes perform pre-processing, including time synchronization and calibration, as well as front-end SLAM for keyframe extraction. To align these independent trajectories, the system relies on cross-view data association, such as feature or map matching, and back-end joint pose graph or factor graph optimization. This tightly coupled fusion not only builds a globally consistent map, but also directly supports downstream exploration, path planning, and navigation tasks.

Recent multi-robot datasets, such as GRACO, provide synchronized multi-modal aerial/ground data with precise calibration for GPS-available and GPS-denied validation~\cite{Zhu2023GRACO}. New air--ground perception benchmarks and methods further indicate that fusing aerial BEV observations with ground views can substantially improve detection and tracking under occlusion. However, communication cost and UAV altitude remain critical design factors: intermediate fusion often achieves a favorable accuracy--overhead trade-off, while perception performance may peak at moderate altitudes, e.g., around $\sim$25\,m, due to scale and depth-estimation limits~\cite{Fang2025pacp}. Large-scale real-world aerial--ground datasets and unified benchmarks are also emerging to quantify these trade-offs more systematically~\cite{Wang2025griffin}. Beyond passive data collection, recent survey work on intelligent active UAV surveillance and monitoring shows that future UAV systems are evolving toward active, predictive, and autonomous sensing agents. This trend is highly consistent with the AG-SCCSI view, where air--ground sensing should be embedded into a closed loop with communication, computing, and decision making~\cite{ahmad2026uavsurveillance}.

With rapidly growing aerial spectrum use, air--ground collaborative spectrum situation awareness also becomes critical for interference mitigation, dynamic spectrum access (DSA), and electromagnetic security. Recent work on 3D radio-environment mapping (REM) leverages sparse Bayesian methods to construct site-specific 3D maps with shadowing information, enabling interferer localization and policy optimization~\cite{TWC2024_3DREM}. UAV-borne sensors can further provide angular diversity for 3D anomaly localization, while multi-angle ``soft--then--hard'' fusion improves detection confidence when single-sensor coverage is limited. Such REM/DT coupling can support angle-dependent access by admitting secondary users in non-interfering angular sectors relative to primary users, thereby improving spectrum utilization without violating protection constraints.

\begin{figure*}[tb!]
    \centering
    \includegraphics[width=0.825\textwidth]{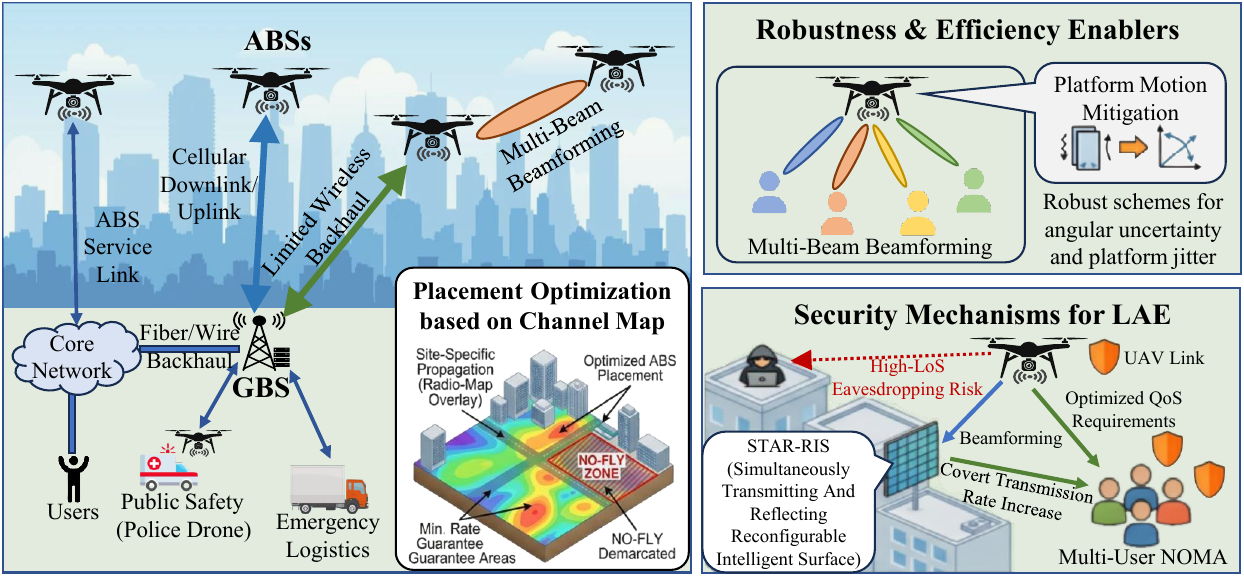}
    \caption{\small{Air-Ground Communication Coordination: Coverage \& Security} }
    \label{fig:communication}
\end{figure*}

\subsection{Orchestrating Air--Ground Communication}
UAVs and HAPs can act as aerial BSs (ABSs)~\footnote{It is debatable whether UAVs can be used as BSs due to the weight addition and potential downlink interference to existing cellular services. Management hassle may prevent this option from practical deployment if cellular systems are considered in LAE.} or relays for coverage restoration/boosting for some LAE applications. As illustrated in the left panel of Fig.~\ref{fig:communication}, this creates a joint air--ground deployment architecture where ground base stations (GBSs) and ABSs coordinate under limited wireless backhaul constraints. A key practical challenge is channel- and constraint-aware 3D placement of multiple ABSs in cities with limited backhaul and no-fly zones. Channel-map (radio-map)--based ABS placement has been shown to guarantee minimum rates to all users by exploiting site-specific propagation instead of distance-only models \cite{Romero2024Aerial}. Beamforming with multi-antenna UAVs improves spectral/energy efficiency but is sensitive to platform motion; robust schemes that account for angular uncertainty and platform jitter (see Fig.~\ref{fig:communication}, top right) are recommended in practice. Finally, secure LAE communications must consider high-LoS eavesdropping risk. Covert/secret transmissions with cooperative NOMA and optimization of active/passive beamforming, including STAR-RIS aiding UAV links, are promising to increase covert transmission rates under wardens \cite{Xu2025multi,Wang2025STAR}. 

ABS deployment should jointly optimize horizontal/vertical positions, backhaul feasibility, and interference coupling. As depicted in the bottom-center panel of Fig.~\ref{fig:communication}, radio-map--driven multi-ABS placement in real maps has matured into tractable convex-relaxation solvers with linear scaling in user count \cite{Romero2024Aerial}. For LAE's sensitive traffic (e.g., for public safety and healthcare logistics), recent works explore collaborative secret--covert access (multi-user NOMA) and RIS/STAR-RIS--assisted covert UAV links (Fig.~\ref{fig:communication}, bottom right) to suppress detectability while meeting QoS requirements \cite{Xu2025multi,Wang2025STAR}. Beyond link-level security, mission-driven orchestration is essential for sophisticated LAE tasks. In our recent work~\cite{ma2026birdcast}, we propose an interest-aware BEV multicasting for infrastructure-assisted collaborative perception. By accounting for individual maps of interest, we formulate a joint feature selection and multicast grouping problem to maximize network-wide utility under communication constraints.

\begin{figure*}[tb!]
    \centering
    \includegraphics[width=0.825\textwidth]{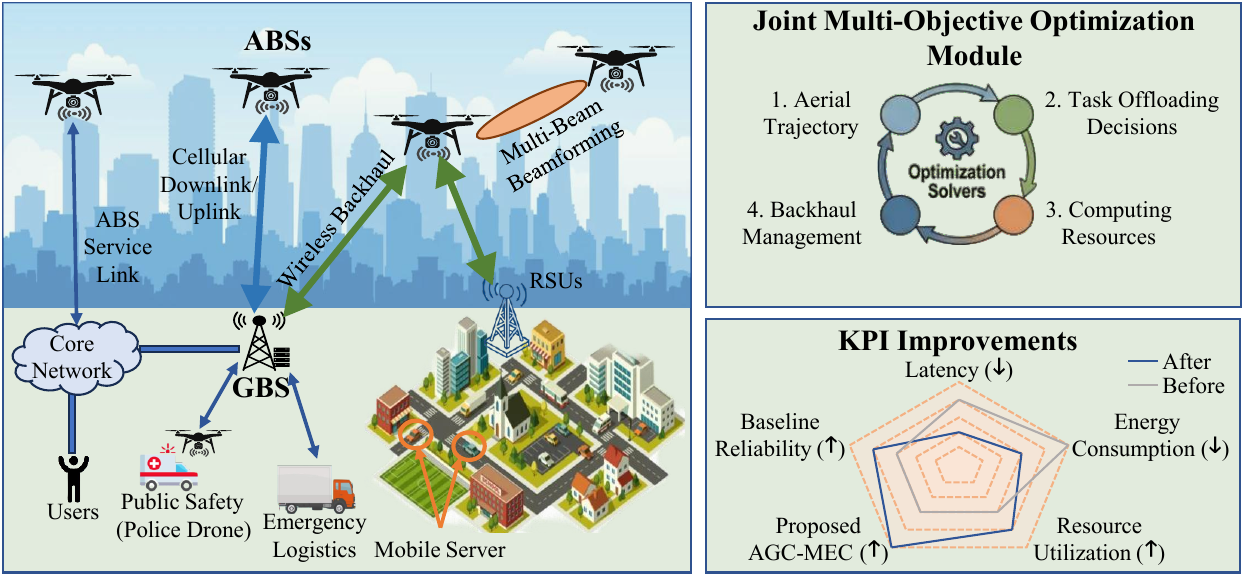}
    \caption{\small{Air-Ground Computing Coordination: multi-access edge computing continuum} }
    \label{fig:computing}
\end{figure*}

\subsection{Orchestrating Air--Ground Computing}
Multi-access edge computing (MEC) offloading cuts down latency and energy consumption; however, terrestrial-only deployments often fail to catch up with the spatio-temporal dynamics of LAE. To address this, an \emph{air--ground collaborative MEC (AGC-MEC)} architecture, as illustrated in the left panel of Fig.~\ref{fig:computing}, unifies aerial servers (e.g., UAVs, airships, balloons) and terrestrial servers (e.g., base stations, APs, RSUs, and ground vehicles) into a cohesive edge computing continuum. Compared with single-domain MEC, AGC-MEC is both more flexible and more powerful for supporting dynamic edge intelligence and task offloading \cite{Zhen2024Air-ground}. 

However, orchestrating such an air--ground synergy requires solving complex resource allocation problems~\cite{deng2026UAV}. As depicted in the right panels of Fig.~\ref{fig:computing}, recent works further address joint multi-objective optimization, encompassing UAV trajectory planning, task offloading decisions, computing resource allocation, and backhaul management. Moreover, integrating joint caching--offloading strategies with reliability constraints in UAV-enabled MEC has demonstrated significant reduction in latency and energy for realistic traffic scenarios \cite{Sun2024TMC,Romero2024Aerial}.

\begin{figure*}[tb!]
    \centering
    \includegraphics[width=0.825\textwidth]{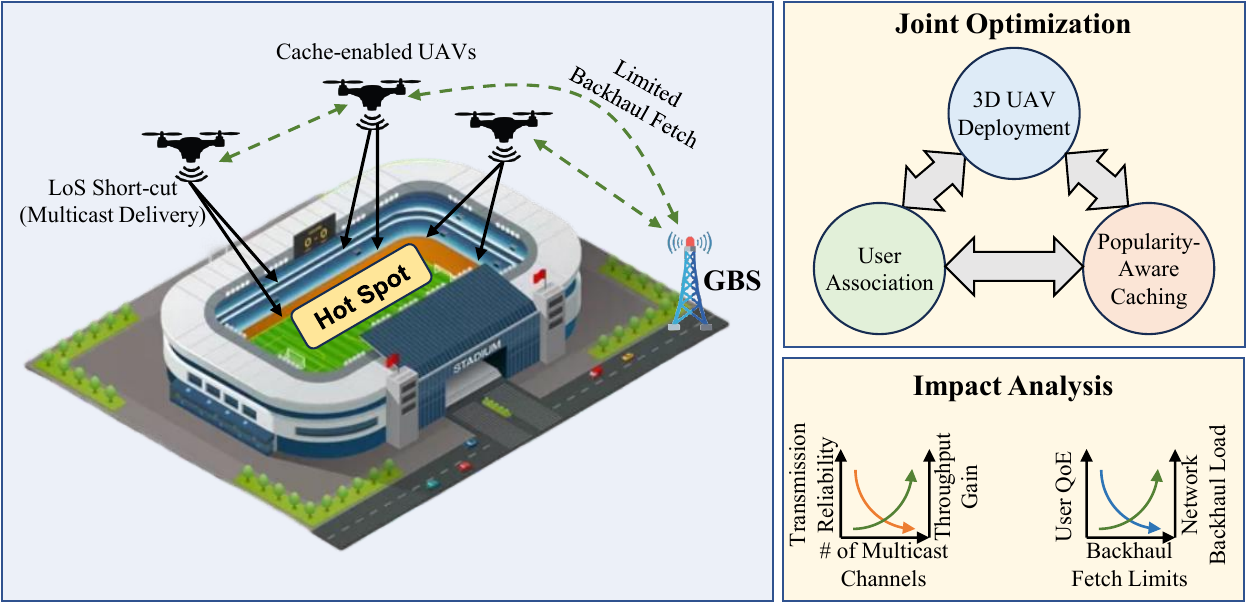}
    \caption{\small{Air-Ground Storage Coordination: edge caching} }
    \label{fig:storage}
\end{figure*}

\subsection{Orchestrating Air--Ground Storage}
Edge caching reduces backbone and backhaul traffic load while minimizing tail latency for recurrent popular contents. UAVs, thanks to their low cost and high mobility, can serve as cache-enabled aerial nodes for temporary hot spots (e.g., football stadiums), significantly improving quality of experience (QoE) via line-of-sight (LoS) short-cuts. As illustrated in the left panel of Fig.~\ref{fig:storage}, orchestrating this air--ground storage involves a strategic split between locally cached content delivery and remote backhaul fetching.

Large-scale analyses that include interference show that popularity-aware (PA) multicast scheduling in cache-enabled UAV networks increases the number of successfully served users. As depicted in the right panels of Fig.~\ref{fig:storage}, optimizing the numbers of multicast channels and backhaul fetched limits is the key to the design under high signal-to-interference ratio (SIR) thresholds \cite{Guo2024Design}. Furthermore, joint 3D UAV deployment, user association, and caching design collectively form a tightly coupled optimization problem, which further reduces the average access delay in multi-UAV cache-enabled networks~\cite{Zhang2025A}.

\subsection{Orchestrating Air--Ground Intelligence}
\begin{figure*}[tb!]
    \centering
    \includegraphics[width=0.825\textwidth]{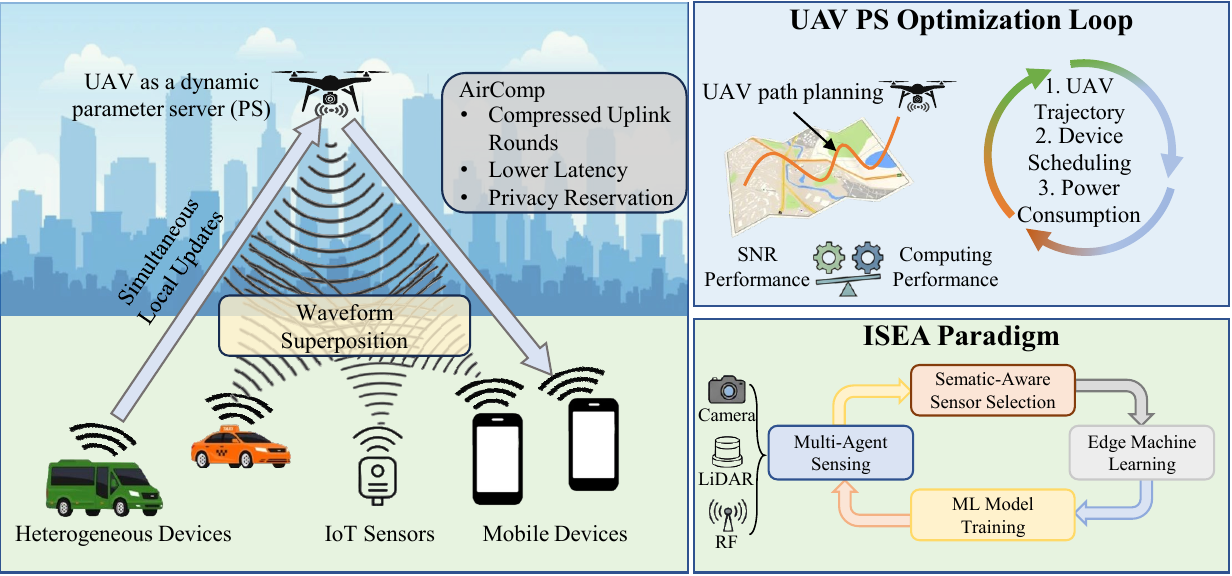}
    \caption{\small{Air-Ground Intelligence Coordination: AirComp-based Federated Learning} }
    \label{fig:intelligence}
\end{figure*}

To navigate over the air safely, the LAE requires on-site learning while preserving privacy. Federated learning (FL) avoids raw data sharing for privacy preservation, but it often suffers from uplink upload overhead and stragglers (links or computing nodes) for model updates. As illustrated in the left panel of Fig.~\ref{fig:intelligence}, \emph{OTA computation} (AirComp) leverages waveform superposition to enable efficient uplink aggregation, while semi-/asynchronous AirComp-FL further mitigates stragglers. Recent results show lower latency and improved round efficiency compared to orthogonal schemes \cite{Zheng2025SAirFed,OTAFLsurvey2024}.

Furthermore, UAVs can serve as agile \emph{parameter servers}. As depicted in the top-right panel of Fig.~\ref{fig:intelligence}, these aerial servers can be dynamically repositioned to balance communication and computing performance, thereby improving fairness and convergence. Energy-efficient UAV-FL jointly optimizes the UAV trajectory, device scheduling, and power consumption~\cite{Fu2025Energy}. Beyond pure FL, the integrated sensing--edge-AI (ISEA) paradigm tightens the loop between sensing and edge ML (see Fig.~\ref{fig:intelligence}, bottom right). Several existing surveys and papers have already identified AirComp-based feature fusion and semantic-relevance-aware sensor selection as key enablers for scalable multi-agent perception \cite{Jiang2025ISEA, AirFusion2024,SemRel2025}.

% \begin{figure*}[htbp]
%     \centering
%     \begin{subfigure}[b]{0.4
%     \textwidth}
%         \centering
%         \includegraphics[width=\textwidth]{figures/collabrations_a.pdf}
%         \caption{Vertical and horizontal collaborations.}
%         \label{fig:sub1}
%     \end{subfigure}
%     \hspace{12pt}
%     \begin{subfigure}[b]{0.4\textwidth}
%         \centering
%         \includegraphics[width=\textwidth]{figures/collabrations_b.pdf}
%         \caption{Typical use cases.}
%         \label{fig:sub2}
%     \end{subfigure}
%     \caption{Illustration for collaborations in AG-SCCSI.}
%     \label{fig:collaborations}
% \end{figure*}

\section{Enabling Technologies for Integrated AG-SCCSI}
\label{sec:enabling}

While Section~\ref{section: framework} establishes the AG-SCCSI framework and identifies the main resource orchestration mechanisms across SCCSI resources, this section focuses on the enabling technologies that make such orchestration practically realizable in LAE. Our emphasis is not on isolated techniques, but on \emph{cross-layer {\rm and} cross-domain enablers} that jointly shape the dynamic behavior of SCCSI resources over air and ground nodes.

\begin{figure*}[t]
    \centering
    \includegraphics[width=0.875\textwidth]{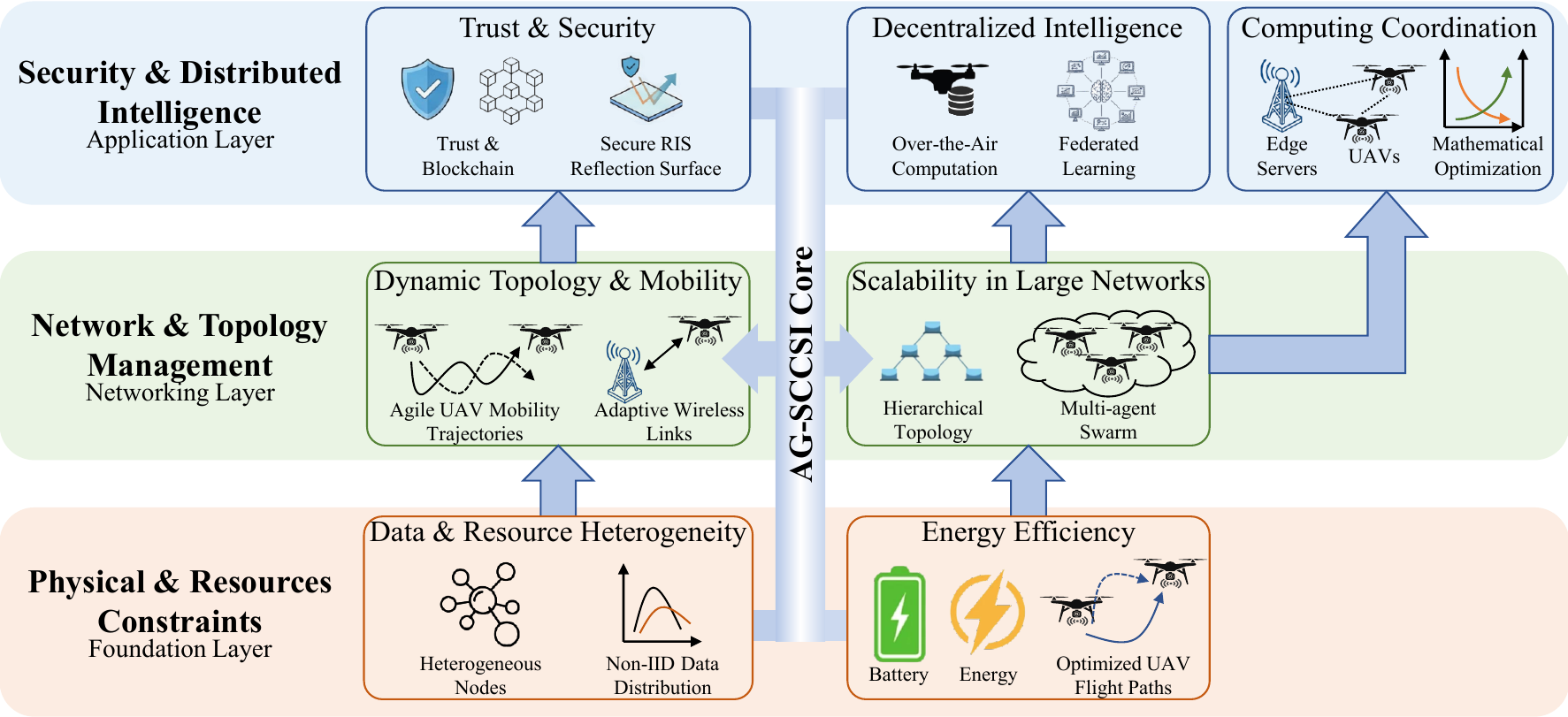}
    \caption{\small{A multi-layered taxonomy of enabling technologies for the AG-SCCSI framework, illustrating the synergistic evolution from foundational resource management and dynamic networking to secure, decentralized intelligence in LAE.} }
    \label{fig:taxonomy}
\end{figure*}

\begin{table}[t!] % 去掉星号，使用单栏标准环境
\caption{Taxonomy of enabling technologies for mission-centric AG-SCCSI in low-altitude economy.}
\label{tab:taxonomy_agsccsi}
\centering
\footnotesize % 恢复到正常的脚注字号，保证可读性
\setlength{\tabcolsep}{3pt}
\renewcommand{\arraystretch}{1.2} % 稍微拉开一点行距

% --- 魔法排版 ---
% 使用 6 个自适应宽度的 X 列，外加 2 个固定宽度的 p 列。
% 6 个 X 列的系数相加必须等于 6 (0.8 + 0.8 + 1.2 + 1.1 + 0.8 + 1.3 = 6.0)
\begin{tabularx}{\textwidth}{@{} 
  >{\hsize=0.8\hsize\raggedright\arraybackslash}X 
  >{\hsize=0.8\hsize\raggedright\arraybackslash}X 
  >{\hsize=1.2\hsize\raggedright\arraybackslash}X 
  >{\raggedright\arraybackslash}p{1.3cm} 
  >{\hsize=1.1\hsize\raggedright\arraybackslash}X 
  >{\hsize=0.8\hsize\raggedright\arraybackslash}X 
  >{\hsize=1.3\hsize\raggedright\arraybackslash}X 
  >{\raggedright\arraybackslash}p{1.2cm} @{}}
\toprule
\textbf{Enabling Pillar} & \textbf{Main SCCSI Dimensions Affected} & \textbf{Primary Control / Design Variables} & \textbf{Typical Timescale} & \textbf{Representative LAE Tasks} & \textbf{Main Coupling With Other Pillars} & \textbf{Main Limitations / Open Issues} & \textbf{Representative Refs.} \\
\midrule

Integrated Sensing and Communication (ISAC) 
& Sensing, Communication 
& Waveform design, beamforming, bandwidth split, sensing--communication resource sharing 
& Fast to medium 
& Obstacle awareness, navigation, spectrum sensing, environment perception 
& Strongly coupled with RIS, DT, SemCom, and FL 
& Sensing--communication tradeoff, mobility robustness, hardware constraints, waveform complexity 
& [2], [51], [54], [56] \\
\addlinespace % 在行与行之间增加一点非常细微的视觉间隔（来自 booktabs 包）

Radio-Map / Digital-Twin (DT) Driven Orchestration 
& Sensing, Communication, Intelligence 
& Radio-map construction, state prediction, what-if evaluation, topology planning, intent-driven control 
& Medium to long 
& ABS deployment, interference-aware access, mission planning, predictive orchestration 
& Coupled with ISAC, AG-MEC, RIS, and application-level scheduling 
& DT fidelity, model update cost, domain gap, online synchronization, scalability 
& [28], [29], [34], [35] \\
\addlinespace

Air-Ground Mobile Edge Computing (AG-MEC) 
& Computing, Communication, Storage 
& Task offloading, server selection, trajectory planning, compute allocation, backhaul management 
& Medium 
& Edge inference, low-latency task execution, collaborative navigation, emergency processing 
& Coupled with caching, DT, SemCom, and FL 
& Joint optimization complexity, volatile connectivity, heterogeneous compute resources 
& [23], [38], [39] \\
\addlinespace

Federated / Over-the-Air Learning (FL / AirComp) 
& Intelligence, Communication, Computing 
& Client selection, aggregation scheduling, trajectory adaptation, OTA aggregation, model placement 
& Medium to long 
& Distributed perception, on-site learning, collaborative model updating, privacy-preserving intelligence 
& Coupled with ISAC, AG-MEC, SemCom, and trust mechanisms 
& Non-IID data, stragglers, communication overhead, trust/privacy risks 
& [42], [43], [44], [48], [49], [52], [53] \\
\addlinespace

Semantic Communication (SemCom) 
& Communication, Intelligence, Storage 
& Semantic extraction, relevance-aware compression, task-oriented source coding, feature transmission 
& Fast to medium 
& Visual navigation, mission reporting, feature delivery under tight uplink constraints 
& Coupled with AG-MEC, FL, and ISAC 
& Task dependence, semantic distortion measurement, cross-model compatibility, benchmark scarcity 
& [13], [14], [31], [32], [45]--[47] \\
\addlinespace

RIS / STAR-RIS 
& Communication, Sensing, Security 
& Phase-shift control, environment shaping, passive beamforming, propagation reconfiguration 
& Fast to medium 
& Coverage enhancement, covert transmission, mobility-resilient links, secure ISAC 
& Coupled with ISAC, DT, and communication planning 
& CSI acquisition, hardware control overhead, deployment practicality in LAE 
& [37], [54], [63]--[66] \\
\bottomrule
\end{tabularx}
\end{table}

As illustrated by the multi-layered taxonomy in Fig.~\ref{fig:taxonomy} and the six-pillar summary in Table~\ref{tab:taxonomy_agsccsi}, several recurring technology families play central roles in AG-SCCSI, including integrated sensing and communication (ISAC), radio-map/digital-twin (DT) driven orchestration, air-ground mobile edge computing (AG-MEC), federated/OTA learning (FL/AirComp), semantic communication (SemCom), and reconfigurable intelligent surfaces (RIS/STAR-RIS). These pillars provide a complementary technology taxonomy. However, because practical LAE missions should be accomplished under multiple intertwined system constraints, the discussion below is organized along seven \emph{cross-cutting challenges}: heterogeneity, topology dynamics, energy-performance tradeoffs, trust and security, scalability, decentralized intelligence, and large-scale computing coordination. In practice, each challenge is typically addressed by multiple technology pillars operating jointly in a closed loop.

\subsection{Addressing Data and Resource Heterogeneity through Cross-Layer Integration}
\label{subsec:heterogeneity}

Heterogeneity is one of the most fundamental obstacles in AG-SCCSI. Air and ground nodes differ significantly in sensing modality, computing capability, energy budget, mobility pattern, communication quality, and data distribution. In LAE, such heterogeneity is further amplified by mission diversity and topology volatility, which makes collaboration substantially more difficult than in conventional terrestrial systems. As a result, heterogeneity cannot be handled at a single layer alone; instead, it requires cross-layer coordination across sensing, communication, computing, and learning.

\subsubsection{Mitigating Sensing-Induced Data Heterogeneity in FL}

In UAV-assisted federated learning, dynamic flight paths and non-uniform sensing conditions lead to highly uneven data acquisition, thereby aggravating non-IID data distributions across clients. Tang \textit{et al.}~\cite{Tang2025ISCC} propose an ISAC-enabled framework in which UAV sensing outcomes directly influence the quality and diversity of collected training data. By introducing a probabilistic sensing model that links UAV deployment to data collection success, they show that sensing-side randomness can significantly affect FL convergence and that such effects can be mitigated through joint optimization of UAV deployment and bandwidth allocation. This result is important for AG-SCCSI because it shows that data heterogeneity in LAE is not only a statistical learning issue, but also a physical-layer and mobility-induced problem.

More broadly, the challenges of statistical, system, and communication heterogeneity in aerial federated learning are surveyed in~\cite{Pham2022Aerial}, which discusses how limited battery, unstable links, and unequal sensing opportunities further worsen imbalance across participating UAVs. The broader lesson is that FL design in LAE should not treat data heterogeneity as a fixed input; rather, it should actively reduce heterogeneity through sensing-aware trajectory design, adaptive communication scheduling, and resource-aware participation control.

\subsubsection{Unified Waveform Design for Balanced ISAC Performance}

Heterogeneity also appears at the functional level: some nodes may be communication-capable but sensing-weak, while others suffer from limited waveform flexibility or hardware capability. Traditional communication-centric signals such as OFDM are often poorly matched to sensing requirements because of strong sidelobes and imbalanced sensing-communication performance. Li \textit{et al.}~\cite{Li2025SLP} address this issue through symbol-level precoding (SLP) for MIMO-OFDM ISAC waveform design, suppressing range-Doppler sidelobes while preserving communication QoS. Their work shows that waveform-level co-design can reduce functional disparity across heterogeneous nodes and enable even lightweight aerial platforms to contribute meaningfully to both sensing and communication tasks.

\subsubsection{Hierarchical Aggregation to Alleviate Non-IID Effects}

When node capability and data quality vary substantially, architectural designs are often needed in addition to physical-layer or scheduling optimizations. Li \textit{et al.}~\cite{Li2025Exploring} propose a hierarchical FL framework in which local groups first aggregate models to reduce intra-group non-IID effects, after which a higher-level aggregator performs global fusion. Such a two-tier design improves robustness against weak or poorly performing clients and reduces the sensitivity of global learning to node-level disparities. In AG-SCCSI, this type of hierarchical architecture is particularly useful because it aligns naturally with air-ground multi-tier resource structures, where UAV clusters, roadside infrastructure, and edge servers may play different aggregation roles.

Overall, heterogeneity in AG-SCCSI should be treated as a coupled sensing-communication-computing-learning problem. Physical-layer design, client scheduling, and hierarchical learning architectures are therefore complementary rather than competing.

\subsection{Managing Dynamic Topologies via Adaptive Sensing and Mobility Control}
\label{subsec:dynamics}

Unlike static terrestrial systems, AG-SCCSI operates over highly dynamic topologies driven by UAV mobility, intermittent links, and rapidly changing mission geometry. As a result, topology management in LAE is inseparable from sensing, mobility control, and communication adaptation. This requirement is closely related to UAV path-planning research, where a recent survey paper has reviewed sampling-based, potential-field, bio-inspired, and AI-based planning methods and analyzed their applicability under different environmental and operational constraints~\cite{kumar2026uavpathplanning}. However, AG-SCCSI goes beyond standalone path planning because mobility decisions must be jointly coordinated with sensing quality, link stability, computing availability, and mission-level service requirements. The enabling technologies in this subsection, therefore, focus on how to maintain robust collaboration under time-varying air-ground connectivity.

\subsubsection{Trajectory-Aware Data Acquisition for Reliable FL}

Fu \textit{et al.}~\cite{Fu2025Energy} study UAV-assisted federated learning under dynamic air-ground links by jointly optimizing UAV trajectory, device scheduling, and communication-computing resources. Their formulation derives scheduling constraints from FL convergence analysis so that only sufficiently reliable devices are selected to participate in each round. This highlights a key design principle for AG-SCCSI: in dynamic topologies, learning reliability depends not only on the FL algorithm itself, but also on mobility-aware and/or reliability-aware participation selection.

\subsubsection{Integrated Sensing for Environmental Awareness}

ISAC further supports topology management by enabling UAVs to perceive their environment while communicating. By exploiting reflected communication signals, UAVs can perform real-time obstacle detection, terrain awareness, or environmental monitoring, and then adapt their trajectories accordingly~\cite{Jiang2025integrated}. This creates a closed sensing-mobility loop in which communication infrastructure is not merely a data pipe but also part of the situational-awareness control mechanism.

\subsubsection{RIS-Enhanced Channel Stability}

Dynamic topology also creates unstable channels due to blockage, angular fluctuation, and UAV motion jitter. RIS provides an additional degree of freedom by reconfiguring the propagation environment in response to mobility-induced changes~\cite{Ma2026RAISE}. Chen \textit{et al.}~\cite{Chen2025RIS} jointly optimize waveform and RIS phase shifts to improve link robustness under time-varying conditions. In AG-SCCSI, this makes RIS not only a coverage enhancer but also a \emph{mobility-resilience enabler} that stabilizes communication for sensing feedback, FL model updates, and latency-sensitive coordination.

Taken together, these works show that topology management in LAE should be formulated as a joint sensing-mobility-communication adaptation problem rather than a pure routing or association problem.

\subsection{Balancing Energy Efficiency and System Performance}
\label{subsec:energy-performance}

Energy is a stringent constraint in LAE, as airborne platforms must balance propulsion power with the energy required for sensing, communication, and computing tasks. Consequently, energy optimization in AG-SCCSI transcends simple communication power minimization, necessitating a multi-dimensional trade-off among mobility, sensing fidelity, and mission-level task execution.

\subsubsection{Energy-Efficient Federated Learning via Joint Optimization}

Fu \textit{et al.}~\cite{Fu2025Energy} minimize the total energy consumption of UAV-assisted FL systems by jointly optimizing UAV trajectory, device scheduling, transmission power, computing power, and local convergence speed and inference accuracy. Their results show that significant energy savings can be achieved without sacrificing model quality when mobility and learning are optimized together. This is particularly relevant to AG-SCCSI, where energy-efficient orchestration must account for both service-layer objectives and physical flight constraints.

\subsubsection{Computation-Aware Compression}

Peng \textit{et al.}~\cite{Peng2024Trajectory} incorporate compression ratio into a joint ISCC design problem and optimize trajectory, resource allocation, and data compression using Multi-Agent Proximal Policy Optimization (MAPPO). Their study shows that compression is not merely a joint source-channel coding issue, but a cross-domain control variable that directly affects communication load, computing effort, and energy expenditure. For AG-SCCSI, this provides an important insight: semantic or task-oriented data reduction can be an effective way to improve energy efficiency without simply reducing mission completion quality.

\subsubsection{Resource-Efficient ISAC under Sparse Occupancy}

Mura \textit{et al.}~\cite{Mura2025Resource} investigate OFDM-based ISAC waveform design under partial spectrum occupancy and show that high sensing accuracy can still be maintained with limited spectral resources. Their use of Schatten-$p$ quasi-norm matrix completion improves sensing efficiency while reducing bandwidth burden. This is especially attractive for LAE systems, where aerial nodes often operate under simultaneous energy and spectrum constraints.

Overall, energy-efficient AG-SCCSI should be understood as a multi-dimensional tradeoff among mobility, sensing quality, communication cost, computing load, and mission timeliness. Single-layer energy optimization is therefore insufficient.

\subsection{Ensuring Trust and Security in Collaborative Intelligence}
\label{subsec:security}

No system can become useful and practical without sound security protection, and thus, in this section, we focus on security issues on AG-SCCSI. Recent survey papers on UAV swarm security show that aerial networks face spoofing, jamming, eavesdropping, hijacking, and coordinated cyber-physical attacks, motivating joint protection across communication, control, and learning layers~\cite{wang2024survey}. Because AG-SCCSI tightly couples cyber and physical operations, failures in trust and/or security can directly lead to mission degradation or even safety-critical incidents. In collaborative air-ground intelligence, secure aggregation, trustworthy participation, incentive compatibility, and physical-layer protection are all essential. 

This security perspective is also consistent with significant research efforts on integrated sensing, communication, and computing in emerging cyber-physical systems. However, AG-SCCSI in LAE further requires explicit air-ground orchestration under mobility, uncertainty, and safety-critical mission constraints, which makes its trust and security design fundamentally more coupled and mission-dependent, and hence more challenging~\cite{wang2024integration}.

\subsubsection{Trust-Aware Participation and Secure Execution}

Trust is the fundamental aspect of security to be addressed. Xu~\textit{et al.}~\cite{Xu2025QFLTrust} propose a Bayesian trust assessment mechanism that evaluates a participant's trust through local model quality and peer recommendations. Such trust-aware selection helps filter out malicious or low-quality contributors before aggregation, thereby improving model integrity. For LAE, this is important because participating aerial and ground devices may differ not only in resource capability, but also in trustworthiness under dynamic and challenged environments.

Recent studies further show that trust assessment should explicitly incorporate delay, energy, and data imbalance. For example, Zheng \textit{et al.}~\cite{Zheng2024TrustTinyFL} develop a lightweight trust metric for tiny-data FL in UAV/IoT systems, while Alaya \textit{et al.}~\cite{Alaya2025UAVFL} review broader trust mechanisms including reputation, blockchain logging, and secure aggregation. These works suggest that trust in AG-SCCSI should be treated as a dynamic orchestration variable rather than a static admission rule.

Beyond participant selection, trustworthy AG-SCCSI also requires protection of the execution environment and the broader UAV cyber-physical attack surface. Trusted execution environments have recently been surveyed as a promising hardware-assisted mechanism for protecting UAV confidentiality and integrity, which is particularly relevant when AG-SCCSI pipelines involve safety-critical sensing, learning, and control decisions~\cite{dong2026teeuav}. Recent work on Internet-of-Drones cybersecurity further develops taxonomies of UAV assets, attacks, countermeasures, and proactive threat-modeling approaches, reinforcing the need for AG-SCCSI security designs that jointly protect communication, sensing, control, and intelligence assets~\cite{spyros2026iodcybersecurity}.

\subsubsection{Incentive Mechanisms for Active Participation}

Our AG-SCCSI framework heavily relies on node/user participation. Collaborative intelligence also depends on sustained and trustworthy  participation. In LAE, however, participating devices may incur various privacy leakage, energy expenditure, and opportunity costs, so incentive design becomes essential. Raja \textit{et al.}~\cite{Raja2025An} develop a Stackelberg-game-based scheme in which leaders provide rewards to participating devices and optimize the tradeoff between fairness and performance. Hu \textit{et al.}~\cite{Hu2024ReliableIncentive} further show that in drone collaborative training, participation drops significantly without appropriate reward allocation and auditing. Hewa~\cite{Hewa2025IncentiveFL} reviews incentive-based FL more broadly and emphasizes the importance of accounting for data quality, privacy cost, and device heterogeneity. For AG-SCCSI, these studies indicate that incentive compatibility is not auxiliary; it is central to maintaining scalable and trustworthy collaboration.

\subsubsection{RIS as a Security Enabler}

RIS can also enhance security by actively shaping the propagation environment. Beyond improving link quality, RIS may suppress eavesdroppers, create favorable alternative paths, or support joint secure sensing and communication. Chen \textit{et al.}~\cite{Chen2025RIS} show how RIS can be integrated with ISAC, while Li \textit{et al.}~\cite{li2023secure} study secure transmission through multiple intelligent reflecting surfaces.  Ahmed \textit{et al.}~\cite{Ahmed2025RISPLS} review RIS-enhanced physical-layer security in UAV-assisted networks, and Li \textit{et al.}~\cite{Li2025RISISAC} discuss RIS-assisted PLS in ISAC systems. Cai \textit{et al.}~\cite{Cai2025Secure} further examine spoofing, jamming, and eavesdropping threats in LAE networking. Collectively, these works show that RIS in AG-SCCSI should be viewed not only as a communication enhancer, but also as a security-oriented environment control mechanism. In fact, UAVs can be leveraged to create more alternative secure paths tp proactively circumvent adversaries. 

In summary, secure AG-SCCSI requires coordinated mechanisms across trust management, participation incentive design, secure aggregation, and physical-layer security protection. No single mechanism alone is sufficient in AG-SCCSI framework.

\subsection{Achieving Scalability in Large-Scale Air-Ground Networks}
\label{subsec:scalability}

Scalability becomes a major concern once LAE moves from isolated UAV missions to dense swarms, large urban deployments, and highly coupled multi-agent, or multi-vendor systems. In such settings, centralized optimization quickly becomes computationally expensive and communication-heavy. The following techniques illustrate how scalable coordination can still be achieved.

\subsubsection{Hierarchical FL for Communication Efficiency}

Hierarchical FL is a natural scalability mechanism because it reduces long-range communication and organizes participants into aggregation layers. Li \textit{et al.}~\cite{Li2025Exploring} show that local aggregation within UAV groups can significantly reduce uplink cost and improve robustness against node failure. In large AG-SCCSI systems, such hierarchical structures also map well onto the natural air-ground resource hierarchy, where local UAV clusters and roadside units can perform intermediate coordination before forwarding compressed updates to edge servers.

\subsubsection{MAPPO-Based Decentralized Coordination}

MAPPO provides a promising approach to scalable multi-agent coordination by allowing individual UAVs to optimize actions using local observations while still learning cooperative behaviors. Peng \textit{et al.}~\cite{Peng2024Trajectory} use MAPPO for integrated sensing, communication, and computing optimization, while Wei \textit{et al.}~\cite{Wei2024MAPPOUAV} demonstrate its effectiveness in dynamic target search. These studies indicate that decentralized reinforcement learning can reduce the need for full centralized control in large UAV swarms, though its training stability and interpretability remain important open issues. More broadly, a recent survey paper on reinforcement learning for UAV systems provides a systematic taxonomy of RL-based UAV control, planning, and coordination methods, further supporting the use of decentralized learning for scalable AG-SCCSI orchestration~\cite{chen2026rluav}.

\subsubsection{Model-Driven Deep Learning for Real-Time ISAC}

Scalability is not only about the number of nodes, but also about real-time scalability. Model-driven deep learning addresses this issue by embedding optimization structure into learnable architectures. Jiang \textit{et al.}~\cite{Jiang2025joint} show that optimization for joint RIS-aided waveform and beamforming design can be unfolded into a model-driven learning architecture, achieving near-optimal performance with much lower online complexity. Complementarily, Cho \textit{et al.}~\cite{Cho2024DRLISAC} investigate an aerial RIS-assisted ISAC system and use deep reinforcement learning to jointly adapt transmit beamforming, RIS phase shifts, and aerial trajectory under dynamic conditions. For AG-SCCSI, such approaches are attractive because they preserve part of the interpretability of model-based design while improving real-time responsiveness in large dynamic networks.

Overall, scalable AG-SCCSI will likely require hierarchical architectures, decentralized control, and learning-accelerated optimization to coexist.

\subsection{Enabling Decentralized Intelligence with Privacy and Heterogeneity}
\label{subsec:decentralized_intelligence}

A defining feature of AG-SCCSI is that intelligence is inheritedly distributed across heterogeneous aerial and ground nodes rather than concentrated in a single cloud or edge server. This decentralization is necessary because LAE involves high mobility, intermittent connectivity, privacy-sensitive data, and mission-driven local decision making. At the same time, it introduces new design challenges related to heterogeneity, trust, and cross-domain coordination.

This view is consistent with prior survey paper on multi-access edge computing and machine learning in the Internet of UAV (IoUAV), which highlights that learning, offloading, and mobility management are deeply intertwined in UAV-centric systems~\cite{ning2023mobile}. It is also aligned with a broader survey paper on machine learning in the Internet of Drones (IoD), which shows that UAV intelligence increasingly relies on the tight interplay among sensing, communication, edge computing, and autonomous decision making~\cite{heidari2023machine}. Compared with these broader IoUAV and the IoD perspectives, AG-SCCSI further stresses explicit air-ground collaboration and end-to-end resource orchestration across all SCCSI dimensions, with explicit roles for storage, semantic communication, and mission-driven closed-loop control.

\subsubsection{Federated Learning under Device and Data Heterogeneity}

In air-ground systems, clients differ in sensing modality, computational capability, battery power, and mobility pattern. These differences induce both statistical and system heterogeneity, which can severely affect FL convergence and inference accuracy. Recent studies show that clustered, hierarchical, and personalized FL architectures are particularly relevant in such settings~\cite{Li2025Exploring}. Zhang and Han~\cite{Zhang2020Federated} demonstrate how multi-UAV networks can benefit from federated coordination, while Cheriguene \textit{et al.}~\cite{Cheriguene} propose participant-selection mechanisms that balance model accuracy, UAV energy consumption, and data heterogeneity. The takeaway here is that FL in AG-SCCSI must be \emph{resource-aware {\rm and} heterogeneity-aware} from the outset.

\subsubsection{Trust, Privacy, and Incentive Mechanisms in Decentralized Intelligence}

Decentralized intelligence also raises privacy and trust concerns because model updates may leak information and adversarial participants may poison training. Kapoor and Kumar~\cite{Kapoor2025Federated} review attacks and defenses in federated urban sensing systems, while Wang \textit{et al.}~\cite{Wang2025Energy} discuss energy-efficient UAV-assisted FL under uncertainty. In parallel, blockchain-assisted clustered FL and trust-based participation mechanisms have been proposed to improve both scalability and robustness. For AG-SCCSI, these results suggest that privacy, trust, and incentives should be designed jointly rather than treated as separate add-ons to FL.

\subsubsection{OTA Aggregation and Mobility-Aware Scheduling}

In delay-sensitive air-ground environments, AirComp can substantially reduce FL uplink latency by aggregating simultaneous transmissions directly over the air during the transmission process. Mobility-aware scheduling is then needed to determine when, where, and from which clients model updates should be collected. Recent studies show that OTA aggregation, asynchronous FL, and client scheduling based on link quality and resource budget are all promising for high-mobility systems~\cite{Wang2025Energy,Zhang2025Latency,Raja2025An,Lee2024Federated}. For LAE, this means that learning rounds must be jointly designed with trajectory, communication, and service timing.

\subsubsection{Edge Intelligence Federation and Cross-Domain Orchestration}

Beyond isolated FL rounds, AG-SCCSI requires \emph{cross-domain intelligence federation} across UAVs, vehicles, roadside infrastructure, and edge servers. This implies that SCCSI decisions jointly shape how intelligence is generated and propagated. Quan \textit{et al.}~\cite{Quan2025Federated} highlight this broader cyber-physical perspective, while \cite{bardesfederated} points to the need for collaborative learning across heterogeneous aerial and ground perception agents. In LAE, cross-domain orchestration is particularly important because aerial and ground nodes often have asymmetric latency, energy, and reliability profiles. Thus, decentralized intelligence should be treated as a multi-domain resource orchestration problem, not merely as distributed model training.

In summary, decentralized intelligence in AG-SCCSI requires joint attention to heterogeneity, trust, mobility, and cross-domain orchestration. These aspects are tightly coupled and should be co-designed. 

\subsection{Scaling Computing Coordination in Large, Uncertain Networks}
\label{subsec:computing_coordination}

A prior survey paper on UAV-enabled edge computing has systematically reviewed computation offloading, trajectory design, and resource allocation from a resource-management perspective, providing a useful baseline for understanding aerial computing coordination~\cite{xia2023survey}. In AG-SCCSI, these issues become broader because computing coordination must be jointly considered with sensing, communication, storage, and mission-level control rather than treated as an isolated MEC  computing resource allocation problem. As AG-SCCSI scales up, the accessible computing resource pool is not fixed; it depends on UAV position, altitude, link quality, and demand distribution, which makes conventional centralized optimization increasingly ineffective in large uncertain environments.

Recent work along this line has begun to interpret this problem from a computing-power-network (CPN) perspective. In particular, Sun \textit{et al.}~\cite{cpn_lae_synergy_2025} connect CPNs with LAE intelligent communications and argue that LAE should be viewed as a broader air-ground orchestration problem in which aerial mobility actively reshapes the accessible computing landscape. This viewpoint is highly relevant to AG-SCCSI because it extends the notion of offloading from simple server selection to mobility-dependent computation-network co-design.

\subsubsection{Evolutionary Algorithms for Scalable Optimization}

Han \textit{et al.}~\cite{Han2024Joint} use a swarm-intelligence-based Fireworks Algorithm to jointly optimize user association, 3D deployment, and flight trajectories in large-scale UAV-assisted MEC systems. Evolutionary algorithms are attractive here because they handle multi-objective, non-convex search spaces without requiring strong analytical structure. Their value in AG-SCCSI lies in offering scalable approximate solutions for complex air-ground coordination problems where exact convex reformulation is hard.

\subsubsection{Stochastic Modeling for Dynamic Environments}

When large-scale systems are too complex to model for deterministic optimization alone, stochastic modeling provides a tractable way to capture uncertainty and spatial randomness. Hui \textit{et al.}~\cite{Hui2024UAV} model user locations via stochastic geometry and define successful edge computing probability as a reliability-oriented performance metric. Their results show that UAV altitude and offloading decisions should be jointly optimized because service coverage and reliability are both spatially uncertain. Such models are particularly useful in LAE because they reveal \emph{structural tradeoffs} before one commits to detailed deployment design.

\subsubsection{Altitude-Dependent Computing Power in MEC Systems}

Deng \textit{et al.}~\cite{Deng2024UAV,deng2024uav_wcm,deng2025uav_c,deng2026UAV} show that in relay-based MEC systems, UAV altitude adaptations not only improve link quality but also reach more ground computing resources. Higher altitudes do improve the accessibility to ground computing nodes, but may also degrade communication quality, thereby creating a non-trivial tradeoff between computation availability and transmission performance. By combining stochastic geometry and queueing-theoretic reasoning, these works highlight that computing coordination in LAE is inherently mobility-shaped and should be analyzed at the end-to-end task level rather than only at the link level.  In~\cite{Fang2025}, we apply the information bottleneck theory to develop a task-oriented feature compression scheme to maximize the end-to-end visual inference quality under bandwidth constraints, which provides a use case of this design philosophy. 

\subsubsection{Multi-Objective Optimization for Large-Scale UAV Networks}

Large uncertain systems also require balancing multiple objectives such as delay, energy, throughput, and task success. Baktayan \textit{et al.}~\cite{Baktayan2024SwarmOffload} investigate multi-agent task offloading and resource allocation for UAV swarms using multi-objective optimization. Such approaches are important for AG-SCCSI because they explicitly recognize that no single metric is sufficient in large mission-driven systems. Instead, computing coordination must be designed under coupled service, mobility, and uncertainty constraints.

Overall, scaling computing coordination in AG-SCCSI requires a combination of structural modeling, scalable heuristics, and mobility-aware communication-computing-buffering co-design. This is one of the clearest areas where LAE departs from conventional terrestrial MEC. This insight is also relevant beyond generic task offloading. For UAV visual navigation in GNSS-denied environments, ground computing resources can serve as an expandable computing and inference substrate, enabling lightweight aerial nodes to offload computation-intensive localization and map-assisted refinement while maintaining closed-loop responsiveness~\cite{Fang2025Task}.

\section{Datasets, Testbeds, and Evaluation Methodology}
\subsection{Datasets and Benchmarks}
\begin{table}[t]
\caption{Representative datasets and benchmarks for AG-SCCSI research in low-altitude economy.}
\label{tab:datasets_agsccsi_revised}
\centering
\footnotesize
\setlength{\tabcolsep}{4pt}
\renewcommand{\arraystretch}{1.12}
\begin{tabularx}{\linewidth}{p{2.45cm} p{1.55cm} p{2.25cm} X p{1.65cm}}
\toprule
\textbf{Name} & \textbf{Type} & \textbf{Main Task} & \textbf{Main Notes / Limitations} & \textbf{Refs.} \\
\midrule
GRACO
& Real-world multimodal dataset
& Cooperative localization and mapping
& Provides synchronized LiDAR, stereo images, IMU, and GPS from aerial and ground perspectives, with millisecond-level synchronization and centimetre-level RTK ground truth. It is highly suitable for cooperative SLAM and map fusion, but mainly focuses on localization/mapping rather than the full AG-SCCSI stack.
& \cite{Zhu2023GRACO} \\

Griffin
& Synthetic benchmark dataset
& Aerial-ground cooperative detection and tracking
& A large-scale aerial-ground cooperative perception benchmark with over 250 dynamic scenes, varied UAV altitudes, and unified protocols for communication efficiency, latency tolerance, and altitude adaptability. It is strong for controlled evaluation, but is still simulation-based.
& \cite{Wang2025griffin} \\

AGC-Drive
& Real-world perception dataset
& Aerial-ground cooperative 3D perception
& A large-scale real-world UAV--vehicle collaborative perception dataset with annotated 3D bounding boxes and public benchmark tools. It is valuable for realistic cross-view perception evaluation, especially under time delays and pose errors, but currently emphasizes perception more than networking or computing.
& \cite{agcdrive2025} \\

OpenUAV / UAV-Need-Help
& Platform + benchmark
& Realistic UAV navigation
& Provides a realistic UAV simulation platform and an assistant-guided object-search benchmark, with about 12k trajectories for UAV vision-language navigation. It is useful for task-driven UAV autonomy, but less directly oriented to communication/computing benchmarking.
& \cite{openuav2024} \\

AERPAW RF Sensor Measurements with UAV
& Real-world RF dataset
& 3D UAV localization / RF sensing
& Contains TDOA localization estimates, ground-truth UAV positions, and LOS/NLOS labels under varying altitudes and bandwidths. It is useful for localization, spectrum, and RF sensing studies, but is communication-centric.
& \cite{aerpaw_rf_2024} \\

AERPAW wireless dataset collection
& Dataset collection / survey
& Aerial wireless propagation and networking
& Consolidates public AERPAW datasets covering I/Q samples, spectrum measurements, LoRa, 5G-NSA, RF localization, and propagation scenarios. It is valuable for reproducible aerial-network research, but does not by itself provide mission-level control annotations.
& \cite{aerpaw_wirelessdatasets_2025} \\
\bottomrule
\end{tabularx}
\end{table}

To provide an objective assessment of the AG-SCCSI framework and its corresponding design mechanisms, it is crucial to offer rich datasets, effective testbeds, and a viable methodology. In this section, we review a representative set of datasets and benchmarks. Table~\ref{tab:datasets_agsccsi_revised} groups representative AG-SCCSI data resources into real-world cooperative SLAM/perception datasets, task-oriented UAV navigation benchmarks, and aerial wireless channel measurements. Existing public resources already support important subproblems such as cooperative localization and mapping, aerial-ground 3D perception, realistic UAV navigation, and RF localization or propagation studies. Representative examples include GRACO for cooperative SLAM, Griffin and AGC-Drive for aerial-ground perception, OpenUAV/UAV-Need-Help for realistic UAV navigation, and the emerging AERPAW wireless datasets for aerial connectivity and localization evaluation.

From an AG-SCCSI perspective, however, the main challenge is not only data scarcity but also benchmark fragmentation. Most public resources still emphasize sensing, perception, navigation, or wireless measurements in isolation, mostly for specific purposes, whereas unified benchmarks that jointly expose synchronized sensing, communication, computing, storage, control actions, and mission executions remain rare. This makes it difficult to compare task-driven orchestration methods under consistent assumptions, especially when latency, reliability, mobility, and safety must be evaluated.

\subsection{Testbeds, Digital Twins, and Simulators}

Table~\ref{tab:testbeds_simulators_agsccsi_revised} summarizes representative experimental platforms for AG-SCCSI research, including open field testbeds, digital-twin frameworks, and modular co-simulation engines. AERPAW is particularly valuable for real-flight wireless experimentation, while the O-RAN DT-RAN provides a structured reference for digital-twin-enabled planning, testing, and energy optimization. At the simulator level, CARLA and AirSim, together with CARLA--AirSim co-simulation and ns-3-based networking modules, offer a practical path toward repeatable evaluation of urban mobility, aerial sensing, UAV control, and wireless channel dynamics. To further bridge the reality gap in complex environments, recent efforts such as~\cite{han2025physics} incorporate physics-informed digital twins to achieve high-fidelity modeling for RIS-assisted wireless systems, ensuring that simulated resource orchestration aligns with physical electromagnetic constraints.

Despite this progress, no single open platform yet provides a mature full-stack LAE evaluation loop, spanning flight physics, air-ground sensing, wireless networking, edge computing and learning, storage or caching, and multi-timescale resource orchestration. A promising direction is therefore interoperable co-simulation, in which scenario generation, network emulation, service execution, and policy control share synchronized states and standardized logs for reproducible mission-level benchmarking.

Existing platforms have already provided partial support for sensing, communication, control, or mobility evaluation, but few expose synchronized observability across all SCCSI dimensions. As a result, full-stack validation of mission-driven air-ground orchestration remains an open challenge. Recent O-RAN-enabled LAE studies further show that open interfaces, RAN intelligent controllers, and xApp/rApp-based closed-loop control can provide a practical software architecture for mission-critical LAE orchestration~\cite{abdalla2026oranlae}.

\begin{table}[t]
\caption{Representative testbeds, digital twins, and simulators for AG-SCCSI research in low-altitude economy.}
\label{tab:testbeds_simulators_agsccsi_revised}
\centering
\footnotesize
\setlength{\tabcolsep}{4pt}
\renewcommand{\arraystretch}{1.12}
\begin{tabularx}{\linewidth}{p{2.55cm} p{1.55cm} p{2.15cm} X p{1.55cm}}
\toprule
\textbf{Platform} & \textbf{Type} & \textbf{Main Scenario} & \textbf{Main Notes / Capability} & \textbf{Refs.} \\
\midrule
AERPAW
& Open field testbed
& UAV + advanced wireless experimentation
& An open aerial wireless experimentation platform built around the convergence of 5G and autonomous drones. It is valuable for real-flight validation, RF measurements, and reproducible aerial networking studies, but does not natively provide a full AG-SCCSI software stack.
& \cite{aerpaw_platform} \\

O-RAN DT-RAN 
& Digital-twin framework
& Planning, testing, energy, and site optimization
& Provides a structured DT-RAN reference covering AI/ML training and testing, network testing automation, network planning, network energy saving, and site-specific optimization. It is not LAE-specific, but is highly relevant as a DT template for AG-SCCSI orchestration.
& \cite{oran_dtran_2024} \\

CARLA
& Urban simulator
& Traffic, sensing, and control
& An open-source urban simulator with flexible sensor suites, actor control, multi-client architecture, traffic scenario generation, and ROS integration. It is strong for ground mobility and urban scene generation, but does not natively model detailed UAV flight dynamics.
& \cite{carla_platform} \\

AirSim
& UAV / robotics simulator
& Drone dynamics and autonomy
& An open-source drone simulator that supports physically and visually realistic simulation with PX4 HITL/SITL integration. It is strong for UAV control and perception prototyping, but usually needs to be paired with other tools for realistic urban traffic and network-layer evaluation.
& \cite{airsim_px4} \\

CARLA--AirSim co-simulation
& Co-simulation stack
& Aerial-ground perception benchmark generation
& Demonstrated in recent aerial-ground cooperative perception benchmarking, this stack combines CARLA urban traffic scenes with AirSim-based drone dynamics. It is useful for synchronized aerial-ground scenario generation, but remains primarily perception-oriented unless combined with network simulators.
& \cite{Wang2025griffin,carla_platform,airsim_px4} \\

ns-3
& Network simulator
& Wireless / protocol / propagation evaluation
& A mature open-source discrete-event network simulator for Internet and wireless systems. Recent releases continue to improve 3GPP propagation support, making ns-3 a strong networking layer for AG-SCCSI co-simulation, though it must be integrated with mobility and task-level simulators for full mission evaluation.
& \cite{ns3_47_2026} \\
\bottomrule
\end{tabularx}
\end{table}

\subsection{Evaluation Metrics for Mission-Centric AG-SCCSI}
Tables~\ref{tab:metrics_mission_system} and~\ref{tab:metrics_dimension_specific} summarize representative evaluation metrics for AG-SCCSI. Unlike conventional communication-centric systems, AG-SCCSI should not be evaluated solely by throughput or isolated task completion performance. Since LAE services are mission-driven and closed-loop by nature, future evaluations should jointly report mission fulfillment, latency, reliability, energy, safety, and orchestration overhead, together with dimension-specific performance indicators for SCCSI.

\begin{table*}[t]
\caption{Mission- and system-level metrics for AG-SCCSI in low-altitude economy.}
\label{tab:metrics_mission_system}
\centering
\footnotesize
\setlength{\tabcolsep}{4pt}
\renewcommand{\arraystretch}{1.12}
\begin{tabularx}{\textwidth}{p{3.0cm} p{1.8cm} X X X}
\toprule
\textbf{Metric} & \textbf{Level} & \textbf{What It Measures} & \textbf{Typical Use} & \textbf{Notes} \\
\midrule
Task Completion Probability / Ratio
& Mission-level
& Probability or ratio that a task is completed within mission constraints
& Offloading, collaborative navigation, edge-aerial task execution
& Requires an explicit task model and deadline definition \\

End-to-End Latency
& Mission / system
& Delay from sensing or request generation to final action or decision
& Navigation, emergency response, logistics, edge inference
& Should ideally be decomposed into sensing, transmission, computing, and control latency \\

Reliability / Deadline Satisfaction Probability
& Mission-level
& Probability that service meets deadline or QoS requirement
& Real-time mission assurance and delay-sensitive tasks
& Reliability targets should be defined according to scenario and mission type \\

Energy Consumption / Energy per Mission
& System-level
& Total or normalized energy cost for mission execution
& UAV planning, FL, offloading, hot-spot caching
& It is preferable to separate propulsion energy from SCCSI service energy \\

Safety Violation Probability
& Mission-level
& Probability of violating safety constraints, no-fly zones, or mission-critical limits
& Dense LAE operations, navigation, airspace management
& Requires an explicit safety model and scenario-specific definition \\

Scalability / Orchestration Overhead
& System-level
& Growth of signaling, control, or computation cost with network size
& Swarm coordination, DT-assisted control, hierarchical FL
& Should include both control-plane and data-plane costs \\
\bottomrule
\end{tabularx}
\end{table*}

\begin{table*}[t]
\caption{Dimension-specific metrics for AG-SCCSI for low-altitude economy.}
\label{tab:metrics_dimension_specific}
\centering
\footnotesize
\setlength{\tabcolsep}{4pt}
\renewcommand{\arraystretch}{1.12}
\begin{tabularx}{\textwidth}{p{3.0cm} p{1.8cm} X X X}
\toprule
\textbf{Metric} & \textbf{SCCSI Dimension} & \textbf{What It Measures} & \textbf{Typical Use} & \textbf{Notes} \\
\midrule
Localization / Mapping Accuracy
& Sensing
& Position error, pose estimation quality, and map consistency
& C-SLAM, GNSS-denied navigation, inspection
& Should be interpreted together with latency and computing cost \\

Detection / Tracking Performance
& Sensing / Intelligence
& mAP, tracking accuracy, ID switches, and occlusion robustness
& Traffic monitoring, surveillance, cooperative perception
& Should be paired with communication overhead \\

Semantic Utility / Task-Relevance Score
& Communication / Intelligence
& Usefulness of transmitted semantic features for downstream tasks
& SemCom, edge-aerial feature delivery
& More task-aligned than raw bitrate, but harder to standardize \\

Throughput / Achievable Rate
& Communication
& Raw transmission capability
& Backhaul, feature delivery, ABS coverage
& Still necessary as a supporting metric, but not sufficient alone \\

Coverage Probability / Outage Probability
& Communication
& Likelihood of service coverage or link failure
& Air-ground connectivity, ABS deployment, relay planning
& Needs scenario-dependent interpretation \\

Age of Information (AoI) / Semantic-AoI
& Communication / Intelligence
& Freshness of updates delivered to controllers or learners
& Monitoring, control loops, sensing updates
& Freshness alone does not measure content usefulness \\

Cache Hit Ratio / Backhaul Load Reduction
& Storage
& How much demand is served locally and how much backhaul is saved
& Aerial caching, hot-spot delivery
& Does not fully capture QoE or fairness by itself \\

Learning Accuracy / Convergence Speed
& Intelligence
& Model quality and training efficiency
& FL, OTA-FL, distributed intelligence
& Should be interpreted together with communication and energy costs \\

Privacy / Security Overhead
& Intelligence / Communication
& Cost of secure aggregation, encryption, authentication, or privacy mechanisms
& Secure FL, trusted collaboration, mission-critical communications
& Often omitted in current simulation studies \\
\bottomrule
\end{tabularx}
\end{table*}

In addition to reporting conventional sensing or communication indicators, future AG-SCCSI evaluations should specify whether the benchmark provides synchronized control logs, edge execution traces, wireless context, and safety annotations. Without such cross-layer observability, it remains difficult to determine whether a method truly improves mission-level performance or only optimizes an isolated subsystem.

\section{Emerging Application Scenarios}

Low-altitude UAV deployment enables a broad range of LAE applications, including traffic management, collaborative logistics, emergency response, and GNSS-denied navigation. In this section, we discuss representative application scenarios and highlight how AG-SCCSI enabling technologies can be orchestrated in closed-loop mission workflows. To make the discussion more concrete, Table~\ref{tab:case_overview_agsccsi} summarizes these scenarios from a mission-centric AG-SCCSI perspective, covering their mission objectives, dominant SCCSI dimensions, and representative enabling technologies. This structured view complements the narrative discussion below and illustrates how different LAE services map to different orchestration priorities.

These case studies are also intentionally selected to reflect the bidirectional nature of air--ground collaboration in LAE. While some scenarios highlight how aerial platforms enhance ground-side sensing and service delivery, others emphasize how ground infrastructures actively support airborne perception, navigation, and closed-loop autonomy through communication, computing, storage, and model assistance.

\begin{table}[t]
\caption{Representative mission-centric AG-SCCSI case studies: scenario overview.}
\label{tab:case_overview_agsccsi}
\centering
\footnotesize
\setlength{\tabcolsep}{3pt}
\renewcommand{\arraystretch}{1.12}
\begin{tabularx}{\linewidth}{p{2.6cm} p{3.0cm} p{3.2cm} X}
\toprule
\textbf{Scenario} & \textbf{Mission Objective} & \textbf{Dominant SCCSI Dimensions} & \textbf{Representative Enabling Technologies} \\
\midrule
\makecell[l]{Real-Time\\Air-Ground\\Traffic Control}
& Wide-area traffic monitoring, congestion mitigation, and incident response
& Sensing, Communication, Intelligence, Computing
& ISAC, SemCom, AG-MEC, DT-assisted orchestration, FL/MARL, RIS/STAR-RIS \\

\makecell[l]{Intelligent\\Collaborative\\Logistics}
& Multi-modal parcel delivery, hub coordination, and time-sensitive logistics under no-fly and mobility constraints
& Communication, Computing, Storage, Intelligence, Sensing
& DT/radio-map-driven orchestration, AG-MEC, caching, MARL/FL, ISAC-based obstacle avoidance \\

\makecell[l]{GNSS-Denied\\Air-Ground\\Navigation}
& High-precision and reliable UAV localization/navigation in urban canyons under strict uplink, latency, and energy constraints
& Sensing, Communication, Computing, Intelligence, Storage
& Task-oriented feature compression, edge--aerial collaboration, ground CPN-assisted computing, map/model caching and refinement, DT-assisted localization\\
\bottomrule
\end{tabularx}
\end{table}

\subsection{UAV-assisted Ground Traffic Control}

UAV-assisted AG-SCCSI can provide a new paradigm for real-time traffic management by combining wide-area aerial perception, ground-side microscopic sensing, edge intelligence, and closed-loop control.

In such systems, UAVs offer a macroscopic view of ground traffic networks by conducting wide-area monitoring, while ground vehicles, including Connected Autonomous Vehicles (CAVs), provide microscopic data on local vehicle interactions. For instance, UAVs equipped with high-resolution cameras and LiDAR can capture real-time traffic density, lane occupancy rates, and abnormal events, such as reckless driving, accidents, or illegal parking, across areas spanning $5$--$10$ km in radius~\cite{Murat2024utilizing}. Instead of overwhelming the uplink with raw video streams, UAVs can employ Semantic Communication (SemCom) to extract and transmit only task-relevant features. These aerial insights complement the detailed data collected by ground CAV sensors regarding vehicle acceleration patterns and driver behaviors. Thus, this hybrid sensing architecture facilitates the identification of potential ``hot spots'', improves bottleneck diagnosis efficiency, and refines traffic alerting systems~\cite{zhang2019machine}.

The feedback control loop in such systems can be characterized as follows:
\begin{itemize}
    \item \textit{Sensing Layer}: UAVs and CAVs cooperatively detect traffic oscillations using Integrated Sensing and Communication (ISAC), which allows the simultaneous tracking of vehicle velocities and data transmission without extra spectrum overhead.
    \item \textit{Communication/Computing Layer}: 5G/6G networks deliver semantic data to edge nodes, with UAVs acting as supplementary airborne edge servers via the AG-MEC continuum to process latency-sensitive tasks locally.
    \item \textit{Decision Layer}: Digital Twin (DT)-assisted orchestration constructs a real-time virtual replica of the traffic network, allowing Multi-Agent Reinforcement Learning (MARL) or Federated Learning (FL) models to generate collaborative control strategies, such as variable speed limits, while preserving data privacy across different vehicle fleets.
    \item \textit{Execution Layer}: Controllable vehicles, such as CAVs and buses, act as ``mobile regulators'' to smooth traffic waves, while UAVs broadcast advisories via V2X communications, potentially assisted by RIS/STAR-RIS to guarantee reliable downlink coverage in urban canyons.
\end{itemize}

Compared with conventional traffic control mechanisms, this AG-SCCSI-enabled system can enhance congestion mitigation, incident response, and transportation throughput through real-time information processing and multi-modal coordination.

\subsection{Ground CPN-Assisted OTA Perception for GNSS-Denied Navigation}

Ground support is equally important in AG-SCCSI, especially when aerial nodes face stringent payload, energy, and onboard computing constraints. A representative example is UAV navigation in GNSS-denied urban canyons, where satellite positioning can be severely degraded by blockage and multipath, while purely onboard visual-inertial solutions often suffer from drift accumulation and substantial real-time computation burden. In this scenario, the ground side should not be viewed merely as a communication backhaul, but as a proactive provider of computing, storage, and machine learning model assistance for aerial localization and navigation.

Following the edge--aerial collaboration paradigm, in our prior work~\cite{Fang2025Task}, the UAV performs multi-view image acquisition together with lightweight task-oriented feature extraction and compression, while the ground-side computing power network (CPN) offers an expandable pool of edge resources for computation-intensive inference, map/model-assisted localization refinement, and closed-loop navigation feedback. Such a design transforms onboard perception from an isolated sensing process into a mission-oriented air--ground collaborative service pipeline.

More specifically, the operational workflow can be summarized as follows:
\begin{itemize}
    \item \textit{Sensing and Representation Layer:} The UAV captures multi-view imagery in complex urban canyons and extracts compact task-relevant visual features instead of transmitting raw video streams. This reduces the uplink burden while preserving navigation-critical information. 
    
    \item \textit{Communication and Access Layer:} Based on uplink quality, latency budget, and resource availability, the UAV selects an appropriate ground point of connection and delivers compressed features to the accessible ground CPN resources through air--ground links.
    
    \item \textit{Ground Computing and Storage Layer:} Ground edge servers execute computation-intensive inference and refinement, including feature matching, map-assisted localization, model-based state estimation, and retrieval of cached maps or navigation priors. In this way, the ground side provides not only stronger computing capability but also persistent storage support that lightweight UAVs cannot easily carry onboard.
    
    \item \textit{Closed-Loop Control Layer:} The refined localization and navigation state is fed back to the UAV for trajectory correction, path replanning, and safe motion control, thereby forming a closed sensing--communication--computing--intelligence loop.
\end{itemize}

Compared with purely onboard autonomy, this ground-assisted design can substantially improve localization robustness and navigation reliability under strict energy and onboard computing constraints. More broadly, it demonstrates a complementary direction of AG-SCCSI: not only can aerial platforms enhance ground-side perception and service coverage, but ground infrastructures can also actively empower OTA perception, airborne decision making, and mission execution through computing, storage, and machine learning support.

\subsection{Intelligent Collaborative Logistics}

The integration of UAVs with ground transportation systems forms an efficient multi-modal logistics network capable of addressing diverse service demands~\cite{gao2026sftridelivercooperativeairground}. In this cooperative framework, UAVs can support suburban or urban-edge delivery tasks, while ground vehicles, couriers, and logistics hubs provide complementary capacity, endurance, and accessibility. Such a dual-layer system can mitigate traffic-induced delays, especially for time-sensitive deliveries such as blood products, emergency pharmaceuticals, and diagnostic specimens.

Recent cooperative delivery designs further explore UAV--ground-vehicle coordination, where aerial mobility and ground transportation resources are jointly scheduled to improve delivery timeliness and service flexibility. For example, TriDeliver models cooperative air--ground instant delivery with UAVs, couriers, and crowdsourced ground vehicles, illustrating how heterogeneous delivery agents can be jointly orchestrated in urban logistics~\cite{gao2026sftridelivercooperativeairground}. The key operational workflow of UAV--ground cooperative logistics systems can be summarized as follows:
\begin{itemize}
    \item \textit{Intelligent Order Parsing}: Upon receiving user orders via mobile or web platforms, the system extracts multi-dimensional parameters, including cargo type, weight, and urgency level. These parameters are tagged with priority labels using NLP-based semantic analysis at AG-MEC servers.

    \item \textit{Cross-Modal Task Allocation}: Leveraging Multi-Agent Reinforcement Learning (MARL) and Trust-Aware Federated Learning (FL), orders are dynamically assigned to UAVs, couriers, and ground vehicles. This decentralized intelligence supports resource utilization while avoiding unnecessary exposure of proprietary delivery data among different logistics operators.

    \item \textit{Air-Ground Collaborative Path Optimization}: Based on 3D urban models, the system generates initial paths and uses real-time environmental data to trigger local obstacle avoidance via ISAC. Meanwhile, trajectory-aware computing coordination synchronizes UAV--ground vehicle rendezvous at logistics hubs, ensuring seamless physical and data handovers.
\end{itemize}

\section{Open Challenges and Future Directions}
\label{sec:future_directions}

Although recent studies have advanced individual dimensions of SCCSI for LAE, a number of fundamental challenges remain open before integrated AG-SCCSI can be deployed at scale. In particular, existing solutions are still largely fragmented across layers, resources, and control timescales, whereas practical LAE services require mission-oriented, closed-loop, and safety-aware orchestration over highly dynamic air-ground systems. In what follows, we highlight several promising research directions that, in our view, are particularly important for the next stage of AG-SCCSI development.

\subsection{Mission-Oriented Closed-Loop Orchestration}
\label{subsec:future_closed_loop}

A central open issue is how to transition from resource-level optimization to mission-oriented closed-loop orchestration. Most existing studies optimize one or two dimensions, such as communication throughput, task offloading latency, sensing quality, or inference accuracy, in isolation. However, LAE missions are inherently end-to-end in nature. For example, a delivery, inspection, or emergency-response task is successful only when sensing, communication, computing, storage, and decision making jointly satisfy latency, reliability, safety, and energy constraints.

Future research should therefore develop \emph{mission-level orchestration frameworks} that directly optimize task completion quality rather than surrogate single-layer metrics. This requires unified abstractions that map heterogeneous mission requirements into multi-resource control variables across air and ground nodes. It also calls for hierarchical control architectures in which fast-timescale adaptation handles link fluctuations, mobility, and sensing uncertainty, while slower-timescale planning governs deployment, caching, model placement, and service migration. Another important direction is the design of \emph{online and predictive orchestration} mechanisms that combine digital twins, mobility prediction, and uncertainty-aware optimization to support look-ahead decision making in dynamic 3D airspace.

\subsection{Energy-Sustainable AG-SCCSI Systems}
\label{subsec:future_energy}

Energy remains one of the most stringent bottlenecks in LAE, especially for UAVs carrying sensing, communication, and computing payloads. Existing work often treats propulsion energy, communication energy, and computation energy separately, which is insufficient for AG-SCCSI systems where these components are tightly coupled. For instance, a higher altitude or longer hovering duration may improve sensing coverage or access to more ground resources, but it also increases propulsion cost and may degrade communication quality.

A key future direction is to establish \emph{mission-aware energy models} that jointly characterize mechanical flight energy and SCCSI service energy under realistic mobility, payload, and environment conditions. Based on such models, future systems should support energy-aware task partitioning, adaptive sensing fidelity control, dynamic communication-computing tradeoff optimization, and charging-aware trajectory planning. Another important topic is the integration of \emph{energy replenishment mechanisms}, such as battery swapping, opportunity charging, and vehicle-assisted or station-assisted energy support, into the orchestration loop. Rather than viewing energy merely as a hard constraint, future AG-SCCSI designs should treat it as the most important system state that directly shapes service continuity and mission feasibility~\cite{guo2025integrated}.

\subsection{Interference-Resilient and Spectrum-Aware Operation}
\label{subsec:future_interference}

As LAE scales toward dense and heterogeneous operations, interference management will become substantially more challenging than in conventional terrestrial systems. Air-ground links are affected not only by co-channel interference from neighboring UAVs and ground infrastructure, but also by strong line-of-sight exposure, dynamic topology changes, and spectrum contention across multiple services. In addition, future LAE systems are expected to support integrated sensing and communication, which further couples spectrum usage with sensing quality and situational awareness. These challenges indicate that interference resilience cannot be addressed by communication-layer adaptation alone. Recent LAE-oriented studies on cellular-connected UAVs therefore emphasize the need for joint communication-control co-design to handle coverage holes, interference uncertainty, and spectrum scarcity in dense aerial deployments~\cite{liang2026controlcommLAE}.

Future research should move beyond static interference avoidance and develop \emph{interference-resilient AG-SCCSI architectures} that combine spectrum awareness, environment perception, and adaptive control. Promising directions include radio-map-assisted spectrum access, digital-twin-enabled interference prediction, trajectory-aware interference coordination, and RIS-assisted or beamforming-assisted propagation shaping. It is also important to investigate how interference can be incorporated into task-level decision making, rather than being treated purely as a physical-layer impairment. For example, future systems may jointly optimize sensing fidelity, communication reliability, and end-to-end quality-aware UAV-assisted computation offloading under interference uncertainty, thereby enabling robust mission execution in crowded low-altitude environments.

\subsection{Scalable Cooperative Intelligence under Heterogeneity}
\label{subsec:future_intelligence}

Distributed intelligence is expected to play a foundational role in AG-SCCSI, but current approaches still face major challenges in scalability, robustness, and various heterogeneities. In practical LAE scenarios, aerial and ground nodes differ significantly in sensing modality, data quality, energy budget, mobility pattern, computing capacity, and communication reliability. These factors lead to severe statistical and system heterogeneities, which can significantly degrade the effectiveness of federated learning, distributed inference, and multi-agent coordination.

A major future direction is to build \emph{heterogeneity-aware cooperative intelligence frameworks} for air-ground systems. This includes hierarchical and clustered learning architectures, asynchronous model aggregation, client selection under mobility and resource constraints, and personalized or task-adaptive model updates for heterogeneous agents. Another promising direction is the joint design of learning and networking, where communication scheduling, trajectory control, sensing decisions, and model aggregation are optimized together. For large-scale swarms and multi-domain systems, scalable coordination also requires decentralized decision mechanisms, such as multi-agent reinforcement learning and game-theoretic control, with explicit consideration of convergence stability, communication overhead, and safety constraints. Importantly, future work should not only improve learning accuracy, but also characterize how distributed intelligence affects mission-level utility, timeliness, and reliability.

\subsection{Safety, Security, and Compliance by Design}
\label{subsec:future_safety}

Safety cannot be treated as an afterthought in LAE. Unlike many traditional wireless systems, low-altitude operations directly interact with physical environments, public infrastructure, and human users. As a result, failures in sensing, communication, control, or intelligence may be translated into tangible operational risks. Meanwhile, the integration of air and ground resources also enlarges the attack surface, including spoofing, jamming, model poisoning, malicious data injection, and unauthorized access to mission-critical services.

Future AG-SCCSI systems, therefore, need to adopt \emph{safety, security, and compliance by design}. One important research direction is the development of assurance-aware orchestration mechanisms that explicitly incorporate risk, uncertainty, and constraint satisfaction into decision making. Another is to establish trustworthy pipelines for data collection, model training, inference, and control execution, supported by secure logging, provenance tracking, and auditable coordination. Beyond cyber security, regulatory compliance should also be translated into machine-interpretable operational constraints, such as no-fly zones, privacy rules, airspace restrictions, and emergency response priorities. More broadly, future work should investigate how to combine formal methods, runtime monitoring, and learning-enabled control to provide certifiable and explainable autonomy for AG-SCCSI systems.

\subsection{Benchmarks, Digital Twins, and Reproducible Evaluation}
\label{subsec:future_benchmarks}

Another critical gap lies in evaluation methodology. Existing studies often rely on isolated simulation settings, task-specific assumptions, and non-unified performance metrics, which makes it difficult to compare different AG-SCCSI designs in a meaningful way. Moreover, many important dimensions in LAE, such as safety margin, mission completion reliability, cross-layer overhead, and adaptability to uncertainty, are still insufficiently benchmarked.

Future research should prioritize the development of \emph{shared datasets, interoperable digital twins, and reproducible testbeds} for air-ground collaborative systems. In particular, digital twins should evolve from passive visualization tools to active experimentation platforms that can support what-if analysis, policy validation, and cross-layer co-simulation of mobility, sensing, networking, computing, and control. Equally important is the establishment of \emph{mission-level evaluation metrics}, such as task completion ratio, end-to-end timeliness, energy per mission, safety violation probability, and robustness under topology or channel uncertainty. Standardized benchmarks along these dimensions would significantly improve the comparability, credibility, and practical relevance of future AG-SCCSI research.

\subsection{Toward a Unified AG-SCCSI Research Agenda}
\label{subsec:future_summary}

Overall, the next phase of AG-SCCSI research should move toward a unified agenda in which mission requirements, multi-resource orchestration, trustworthy intelligence, and reproducible evaluation are treated in an integrated manner. The long-term challenge is not merely to connect aerial and terrestrial systems, but to enable them to cooperate as a coherent, adaptive, and verifiable service fabric for LAE. This service fabric may also extend beyond air-ground collaboration toward multi-tier aerial infrastructure. For instance, recent work highlights the potential of high-altitude platforms as supervisory and computing-capable infrastructure for LAE, bridging communication, computing, navigation integrity, and regulatory oversight across regional low-altitude operations~\cite{huang2026haplae}. Addressing these challenges will require advances in system modeling, algorithm design, digital twinning, distributed intelligence, and standardization. We believe that such a shift from fragmented optimization to mission-centric integration will be essential for realizing scalable and trustworthy low-altitude services.

\section{Conclusion}

This survey has presented a mission-centric view of low-altitude economy (LAE) through the unified AG-SCCSI framework, which integrates sensing, communication, computing, storage, and intelligence across air and ground systems. Rather than treating UAV networking, edge computing, caching, or distributed intelligence as isolated design problems, we have argued that practical LAE services must be understood as closed-loop multi-resource orchestration problems shaped by mobility, uncertainty, safety, and heterogeneous mission requirements.
From this perspective, we have first introduced the AG-SCCSI framework and the corresponding air-ground resource orchestration mechanisms. We have then reviewed the enabling technologies that support integrated AG-SCCSI, including collaborative sensing and ISAC, air-ground communications, AG-MEC, storage and caching, decentralized intelligence, and large-scale computing coordination. Beyond technologies alone, we have also summarized representative datasets, testbeds, digital twins, simulators, and mission-centric evaluation metrics, and discussed representative application scenarios to illustrate how different LAE services map to different orchestration priorities.
A central takeaway of this survey is that the main bottleneck in LAE is no longer mere connectivity. Instead, the core challenge lies in how to translate application-driven mission requirements into coordinated sensing, communication, computing, storage, and intelligence decisions over highly dynamic air-ground systems. This requires moving beyond siloed optimization toward unified, trustworthy, and reproducible air-ground service design.
To the best of our knowledge, this is the first survey that systematically frames LAE from the perspective of AG-SCCSI. We hope this work can serve not only as a structured reference on the current literature, but also as a roadmap for future research on mission-oriented orchestration, scalable cooperative intelligence, safety-aware autonomy, and reproducible validation for low-altitude services.

% ---------------------------------------------------------
% 6. 将原先的 \thanks 内容移到论文末尾的 acks 环境中
\begin{acks}
The research work described in this paper was conducted in the JC STEM Lab of Smart City funded by The Hong Kong Jockey Club Charities Trust. This work was also supported in part by a grant from the Research Grants Council of the Hong Kong Special Administrative Region, China (Project No. CityU 11216324) and in part by the Hong Kong SAR Government under the Global STEM Professorship. The work of Y. Deng was also supported in part by the National Natural Science Foundation of China under Grant No. 62301300 and in part by the Shandong Province Science Foundation under Grant No. ZR2023QF053.
\end{acks}

\bibliographystyle{ACM-Reference-Format}
\bibliography{deng}

\end{document}